\documentclass[11pt]{article}
\usepackage[utf8]{inputenc}
\usepackage[margin=1.25in]{geometry}
\usepackage{hyperref}
\usepackage[titletoc,toc,title]{appendix}
\usepackage{amsmath}
\usepackage{cite}
\usepackage{amsfonts}
\usepackage{amssymb}
\usepackage{mathtools}
\usepackage{graphicx} 
\usepackage{graphics}
\usepackage{subcaption}
\usepackage{braket}
\usepackage{xcolor}
\usepackage{authblk}
\usepackage{float}
\usepackage{ulem} 
\usepackage{comment}

\newcommand{\ssee}{{\mathcal{S}}}
\newcommand{\sseec}{{\ssee}^{(c)}}

\newcommand{\mink}{\mathbb M}
\newcommand{\diam}{\mathbb D} 
\newcommand{\cH}{\mathcal H}
\newcommand{\cR}{\mathcal R} 
\newcommand{\mx}{\mathrm{max}}
\newcommand{\mn}{\mathrm{min}}

\newcommand{\Ns}{N_O}
\newcommand{\hD}{\hat \Delta}

\newcommand{\deS}{\mathrm{dS}}
\newcommand{\con}{\alpha}
 
\author[1]{\textbf{Sumati Surya}}
\author[1]{\textbf{Nomaan X}}
\affil[1]{\textit{Raman Research Institute, Sadashivnagar, Bangalore 560 080, India}}
\author[2,3]{\textbf{Yasaman K. Yazdi}}
\affil[2]{\textit{Theoretical Physics Group, Blackett Laboratory, Imperial College London, \newline
SW7 2AZ, UK}}
\affil[3]{\textit{Department of Physics, 4-181 CCIS, University of Alberta, Edmonton AB, \newline
T6G 2E1, Canada}}
\affil[ ]{\textit {ssurya@rri.res.in, nomaan@rri.res.in, ykouchek@imperial.ac.uk}}
\title{{\textbf{Entanglement Entropy  of \\ Causal Set  de Sitter Horizons}}}
\date{}

\begin{document}
\maketitle
\begin{abstract}

{de Sitter cosmological horizons are known to exhibit thermodynamic  properties similar to black hole horizons. In this
work we study causal set  de Sitter horizons, using  Sorkin's spacetime
entanglement entropy (SSEE) formula, for a conformally coupled quantum
scalar field.   We calculate the  causal set SSEE for the Rindler-like wedge of a symmetric slab of de
Sitter spacetime in $d=2,4$ spacetime dimensions using the Sorkin-Johnston vacuum state. 
We find that the SSEE obeys an area law  when the spectrum of the  Pauli-Jordan operator is appropriately  truncated in both the de Sitter
slab as well as its restriction to the Rindler-like wedge. 
Without this truncation,  the SSEE
satisfies a volume law. This is in agreement with Sorkin and Yazdi's calculations for the causal set SSEE for nested
causal diamonds in $\mink^2$, where they showed that an area law is obtained only after truncating the Pauli-Jordan spectrum. 
In this work we explore  different truncation schemes with the criterion that the SSEE so obtained 
obeys an area law. } 
\end{abstract} 


\section{Introduction}

Cosmological horizons {in de Sitter ($\deS$) spacetime} share several  key features with black hole horizons \cite{Bekenstein:1972tm, Bekenstein:1973ur,
  Bekenstein:1974ax}, as first suggested in \cite{gibbons}. Classically, both can be associated with a temperature, {as
well as} 
an entropy proportional to the horizon area based on a mathematical analogy {with} the laws of thermodynamics. Quantum
mechanically, observers outside both horizons can detect thermal radiation characterised by the horizon temperature.
However there are also key differences \cite{Davies:1988dma}. Most obvious is the fact that different observers in de
Sitter have different corresponding horizons. Moreover, the thermality of $\deS$  radiation is not reflected in the
stress energy tensor of the quantum state and is instead red-shifted by the expansion. {Despite this, the entropy-area
relationship is robust and can moreover be extended to all causal horizons \cite{Jacobson:2003wv}.} 

The interaction of matter fields with black hole horizons also exhibits thermodynamic features. As in the case of a
black body, incoming radiation is scattered into thermal radiation at around the black hole temperature
\cite{panangaden}.  In \cite{1983ee} Sorkin proposed that the dominant contribution to black hole entropy  can potentially come from the entanglement entropy (EE) of a non-gravitational
field. This EE was defined using the reduced density matrix of the exterior region. An
explicit calculation for a scalar field was carried out in \cite{Bombelli:1986rw} and seen to give rise to an
area law after imposing a UV cutoff. An area dependence arises naturally from complementarity and is an
important feature of EE. It has been shown to hold for a diverse range of quantum systems \cite{cardy}.

Numerous researchers have since studied the connection between {EE} and black hole entropy \cite{ted, solod, Emparan, Jacobson:2012yt}. In \cite{jacobson1994black} Jacobson suggested that the ``species puzzle'' can be resolved
by showing that the renormalisation of the gravitational constant appearing in the Bekenstein-Hawking entropy is
similarly species dependent. In recent years, the idea of holographic EE has gained considerable ground starting with
the work of Ryu and Takayanagi \cite{Ryu}. The EE in $\deS$  was first calculated in \cite{Maldacena:2012xp} and shown to
exhibit the area law relation, both for a free massive field theory using the $\deS$  Euclidean vacuum, as well as for strongly coupled field theories with holographic duals (see also \cite{Dong:2018cuv}).

All these calculations of the EE use the density matrix specified on a partial Cauchy hypersurface ${\Sigma}$, with the entropy attributed to its spacetime domain of dependence $\mathcal{D}({\Sigma})$.
However, it is desirable to define the EE in a more covariant language, since horizons are intrinsically spacetime in character. In \cite{see} Sorkin proposed a spacetime EE, which we term the Sorkin Spacetime Entanglement Entropy or SSEE for short, defined for a Gaussian free scalar field theory. The SSEE between a globally hyperbolic subregion $O$ in a globally hyperbolic compact\footnote{The compactness condition on $\mathcal{M}$ is important in defining the SSEE, since the domain of the integral operator $i\hat \Delta$ is the space of compactly supported functions, while its range includes functions that are not of compact support.} spacetime region $\mathcal{M}$ and its causal complement is given by  
\begin{equation}
  \ssee=\sum_{\mu} \mu\ln |\mu| \label{s4} 
\end{equation}
where $\mu$ is the generalised eigenvalue 
\begin{equation} 
W_O(x,x') v=i \, \mu \, \Delta_O(x,x') v,\indent \Delta_O v\neq 0,
\label{sw4}
\end{equation}
and $W_O(x,x')$ and $ i\Delta_O(x,x')$ denote the restrictions to $O$ of the Wightman function
$W(x,x')=\langle0|\phi(x)\phi(x')|0\rangle$, and the Pauli-Jordan function $i\Delta(x,x')=[\phi(x),\phi(x')]$, respectively.
Recently this formula has been shown to be valid up to first order in perturbation theory for generic perturbations away from the free field Gaussian theory as well \cite{ngsee}; in this case the Gaussian free field correlation functions are replaced with their perturbation-corrected counterparts. 

\begin{figure}
 	\centering
 	\includegraphics[scale=0.85]{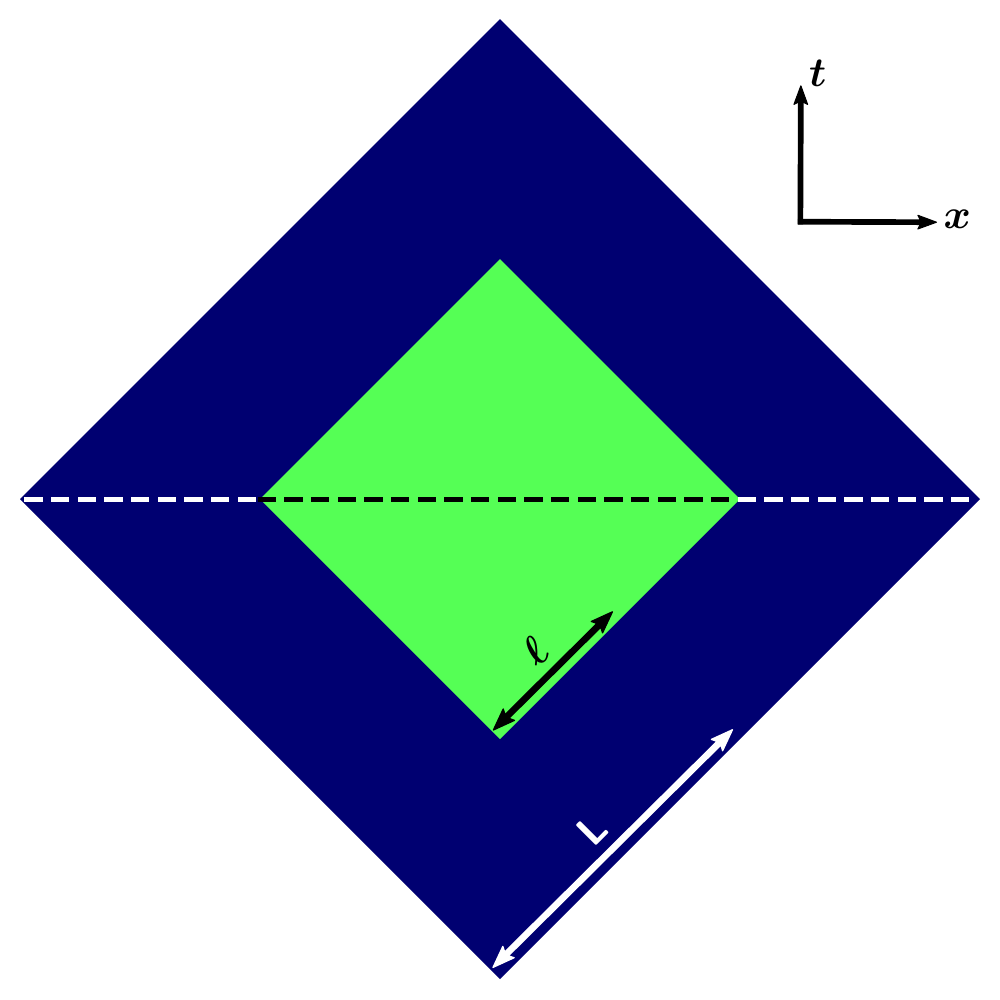}
 	\caption{The causal completion or domain of dependence $\mathcal{D}$ of two line segments, one contained within the other.}
 	\label{diamonds}
 \end{figure}

In \cite{Saravani:2013nwa}  $\ssee$  was calculated for nested  causal diamonds in $d=2$
continuum Minkowski spacetime $\mink^2$, $\diam_\ell^2 \subset \diam_L^2$, which are each the domain of dependence of nested  spatial intervals of lengths $2\sqrt{2}\ell$
and $2\sqrt{2}L$, respectively, as shown in  Figure~\ref{diamonds}. Rather than the Minkowski vacuum, the calculation
of \cite{Saravani:2013nwa} used the covariantly defined Sorkin-Johnston (SJ) vacuum for free scalar fields \cite{Johnston:2009fr,Sorkin:2017fcp}.
 As in other calculations of EE, $\ssee$ can be calculated in the continuum only after imposing a UV cutoff.  The SJ
 vacuum offers the choice of a covariant cutoff in the eigenspectrum of the Pauli-Jordan operator $i\hD$ (the SJ spectrum), which is
 at the heart of the SJ construction. Using this cutoff it was shown in  \cite{Saravani:2013nwa}  that the $\ssee$ satisfies the
 expected $d=2$ ``area'' law.

 Since a causal set which is approximated by a continuum spacetime comes with a built-in covariant spacetime cutoff $1/\rho$,
 one might expect that the  SSEE for a causal set doesn't need further regularisation. While it is finite for a finite causal
 set, it was shown in \cite{Sorkin:2016pbz} that the SSEE in the causal set version of the  calculation in
 \cite{Saravani:2013nwa} obeys a spatial area law only after a suitable ``double truncation'' of the causal set SJ spectrum both in
 $\diam_L^2$ and in $\diam_\ell^2$.  Without this, the SSEE follows a spacetime volume law and thus violates
 complementarity.

\begin{figure}[!h]
	    \centering
	    \includegraphics[width=0.85\textwidth]{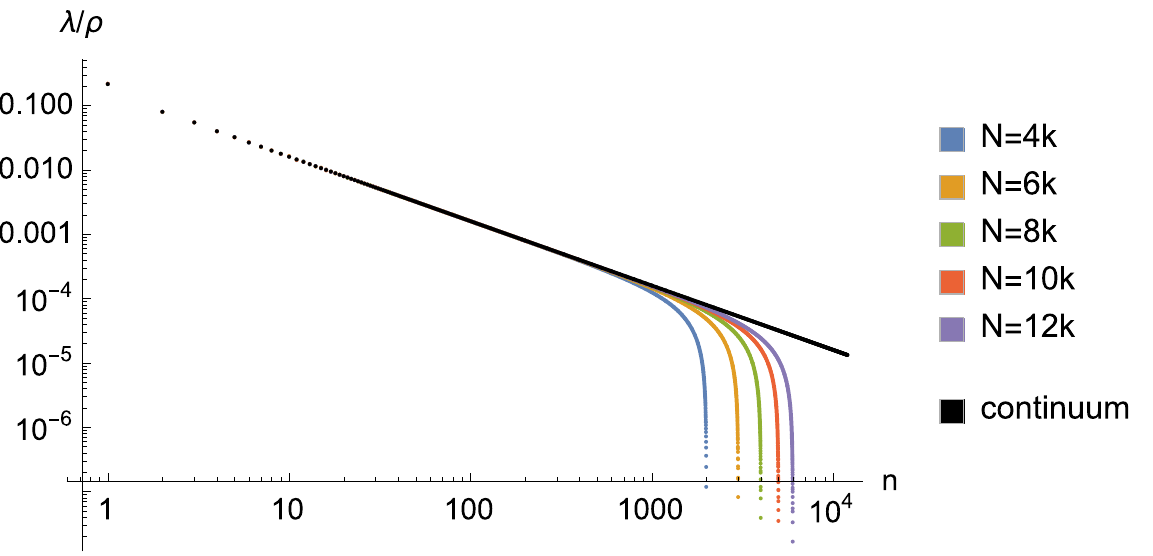}
	    \caption{Log-log plot of the normalised SJ spectrum for the $2d$ causal diamond of side length $2L=1/\sqrt{2}$, for both the continuum as well as for causal sets of size  $N$. On the y-axis, $\lambda$ refers to the SJ eigenvalue (see appendix B) and $1/\rho$ is the covariant volume cut-off coming from the causal set.}
	    \label{fig:knee}
          \end{figure}

 The double truncation used in \cite{Sorkin:2016pbz} was motivated by comparing the SJ spectra of the 
 continuum with that of the causal set in $\diam_L^2$.  The latter possesses a characteristic ``knee" at which the
 eigenvalues dramatically drop to small but non-zero values (see Figure \ref{fig:knee}).  It is roughly around  this knee that the discrete and continuum spectra begin to disagree.  
Importantly, while the formula  for the SSEE  \eqref{s4} excludes {solutions with strictly zero eigenvalues}, it does
not exclude {those with} finite {\it near} zero eigenvalues, which characterise the post-knee {causal set} SJ spectrum. These
modes can be shown to contribute to large $\mu$ values in \eqref{s4} which then dominate the SSEE. If we include eigenfunctions $v$ that lie in the kernel of $i\Delta$ but not necessarily of $W$, this gives an infinite
contribution to $\ssee$, since the equation can only be satisfied for $\mu \rightarrow \infty$ (this is also discussed in 
\cite{Belenchia:2017cex}). Thus, in the causal set,
the contribution from a $v$ which is {\it almost} in the kernel, i.e., $||i \Delta v|| \approx 0$, lends itself to a
very large (though finite) value of $\mu$, and hence to a much larger SSEE.     

Extending this work to gravitational horizons is of course very  important, not least because 
causal sets provide a covariant  UV cutoff, essential to the finiteness of EE.  Calculating the SJ vacuum for a free
scalar field  in the
causal set needs the proper identification of the dimension dependent causal set retarded Green function
\cite{Johnston:2010su}. While such an identification is not yet known for black hole spacetimes, it has been obtained  for
$\deS$   \cite{NomaanX2017}, which in turn gives us the ability to calculate the causal set $\deS$ 
SSEE. The causal set $\deS$  SJ vacuum was obtained in \cite{Surya:2018byh} and found to differ significantly from the
known continuum $\alpha$-vacua. In the continuum it is known that the latter are the only possible $\deS$  invariant
vacua, which suggests that the causal set discretisation and/or the SJ prescription has a non-trivial effect on the QFT vacuum.   
 
In this work we calculate the causal set SSEE for the $\deS$  horizon,   for a conformally coupled, massless, free scalar
field in dimensions $d=2,4$.  We find that, as for nested causal diamonds in $\mink^2$, the SSEE  obeys an area law only
after a suitable double truncation, without which it follows a spacetime volume law.  The truncation scheme used in \cite{Sorkin:2016pbz} used the explicit analytic form of the
SJ spectrum in the $2$d flat spacetime causal diamond to motivate the truncation in the causal set SJ spectrum. The
analytic form of the SJ spectrum is however not known more generally. {In this work we motivate the choice of truncation scheme
  for the $\deS$  causal set SJ spectrum by requiring the causal set SSEE to satisfy  an area law. As we will see, satisfying this criterion is  quite non-trivial.}

Section \ref{prelims} provides a background for  our work. In Section \ref{areas}  {we begin with a discussion of
  area laws and complementarity. We define the
  two complementary  Rindler-like wedges in $\deS$  and the corresponding Bekenstein-Hawking  area law which we} might 
expect to recover from the SSEE. In Section \ref{ssee_in_cs} we set up the calculation of {the} SSEE in {a}  finite causal set. In Section
\ref{2dmink_review} we review the results of the calculation of SSEE  for nested causal diamonds in
$\mink^2$ \cite{Sorkin:2016pbz} {and} the {critical} role {played by the} double truncation {procedure} in obtaining the area law. In
Section \ref{gen_trunc} {we propose generalisations of the truncation scheme of  \cite{Sorkin:2016pbz} for general
  spacetimes},  in the absence of analytic results on the SJ spectrum in the continuum.\footnote{{The 
  continuum SJ spectrum in the $\deS$  slab has been recently obtained \cite{AMSS}, but not  in the Rindler-like
  wedge.}} 

In Section \ref{results} we present the results of extensive numerical simulations {for the causal set SSEE for  
$\deS_{2,4}$ horizons. Our investigations of different truncation schemes show that an area law compatible with the
  Bekenstein-Hawking entropy of the horizon is 
  not easy to satisfy. Complementarity on the other hand is guaranteed, up to Poisson fluctuations,  by the fact that the
  Rindler-like wedges are identical in the continuum.  We present a few truncation schemes and discuss their relative merits.}
We end with  open questions in Section \ref{discussion}. The
Appendices contain some of the background material on the SJ vacuum as well as {a calculation of the causal set SSEE
 for} nested causal diamonds
in $\mink^4$, where {finding} a  truncation that satisfies both an area law and complementarity has proven to be more non-trivial.  

\section{Preliminaries} 
\label{prelims} 
\subsection{Complementary Regions in Global $\deS$  } 
\label{areas}

\begin{figure}[!h]
	    \centering
	    \includegraphics[width=0.6\textwidth]{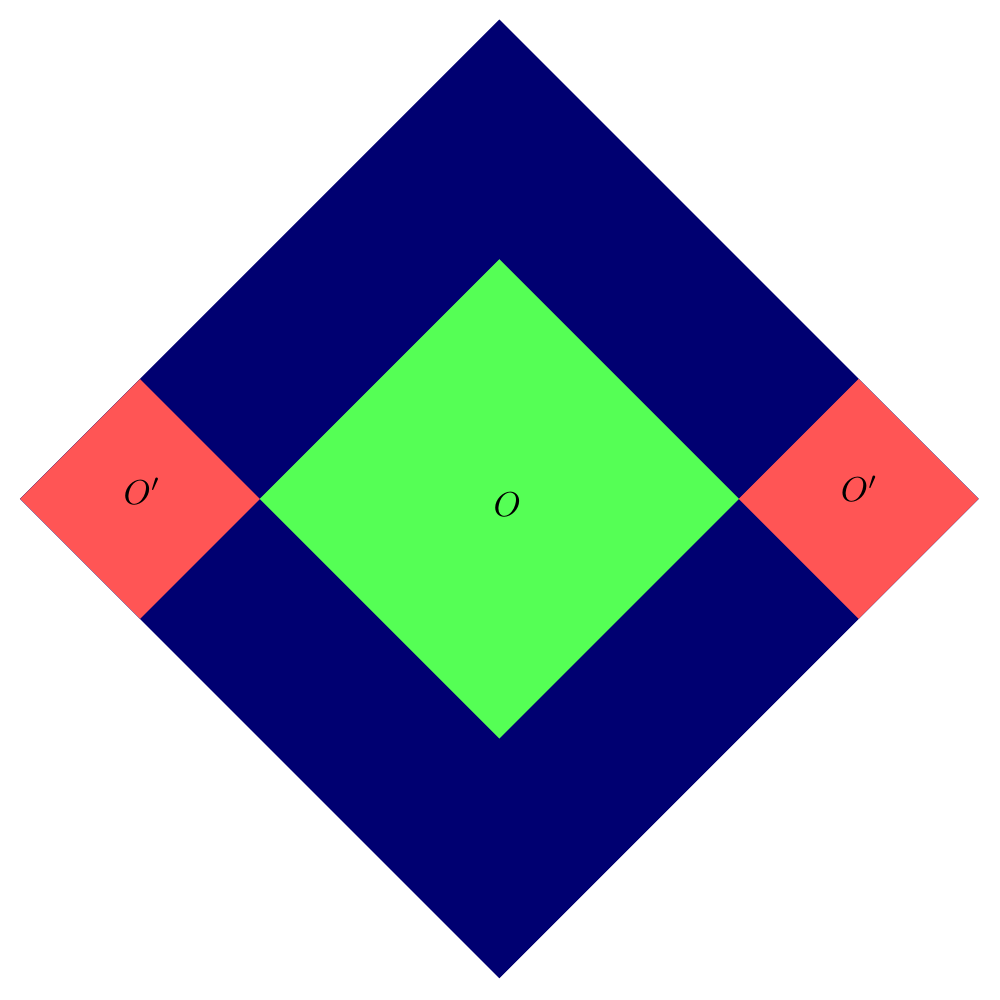}
	    \caption{A nested causal diamond $O$ and its complement $O'$ in $\mink^2$.}
	    \label{fig:2ddiamonds}
          \end{figure}

Let $(\mathcal{M},g)$ be a globally hyperbolic spacetime region and $O$ a globally hyperbolic subregion $O \subset \mathcal{M}$.
The  SSEE of
$O$ is defined with respect to its causal complement $O'$, where
$O' \subset O^c\subset \mathcal{M}$ such that $ \, \, x \in O'
\Leftrightarrow x$ is spacelike to $O$.  Since $(\mathcal{M},g) $ is globally hyperbolic, so is $O'$  and hence
the EE of $O'$ can also be defined with respect to $O$,   which is {\it its} causal complement.
$O$ and $O'$ are said to be {\sl complementary} to each
other, where  we now use the term ``complementarity'' to denote causal complementarity.  Figure \ref{fig:2ddiamonds}
shows an example of a smaller causal diamond $\diam^2_\ell$  nested inside a larger one $\diam_L^2$ in $\mink^2$. The complement $O'$  to $O \sim \diam_\ell^2$ is a union of two disconnected causal diamonds.  
In \cite{Saravani:2013nwa} and \cite{Sorkin:2016pbz} the SSEE of $O$ with respect to $O'$ was calculated in the
continuum and in the causal set, respectively.   Note that in the standard definition  the spatial 
complementary regions $\Sigma_O$ and $ \Sigma_{O'}$ are used to define EE, where $\Sigma_{O}$ denotes a partial Cauchy
hypersurface of the region $O$  in  $\Sigma$, a Cauchy hypersurface of $(\mathcal{M},g)$. However, because $O$ is globally
hyperbolic,  the ``information content'' of $\Sigma_{O}$ is the same as that of $O$.

A feature of bipartite EE is that it satisfies complementarity, i.e.,  that the EE of $O$ with respect to its complement
$O'$ is the same as that of $O'$ with respect to $O$. This in turn implies the area law since the two complementary
regions only share a spatial boundary separating them. The gross feature of this boundary is its  ``area'' or $d-2$
spatial volume, which means that the EE satisfies an area law.\footnote{The EE could also depend on the more detailed
  geometry of the boundary, but we will ignore this possibility in our work. See \cite{Sorkin:1999yj} for a discussion
  on this.} Conversely, a scaling of the EE with the spatial or spacetime volume of the region means that
complementarity is not satisfied, since in general the volumes of $O$ and $O'$  can be unequal.

\begin{figure}[!h]
	    \centering
	    \includegraphics[width=0.58\textwidth]{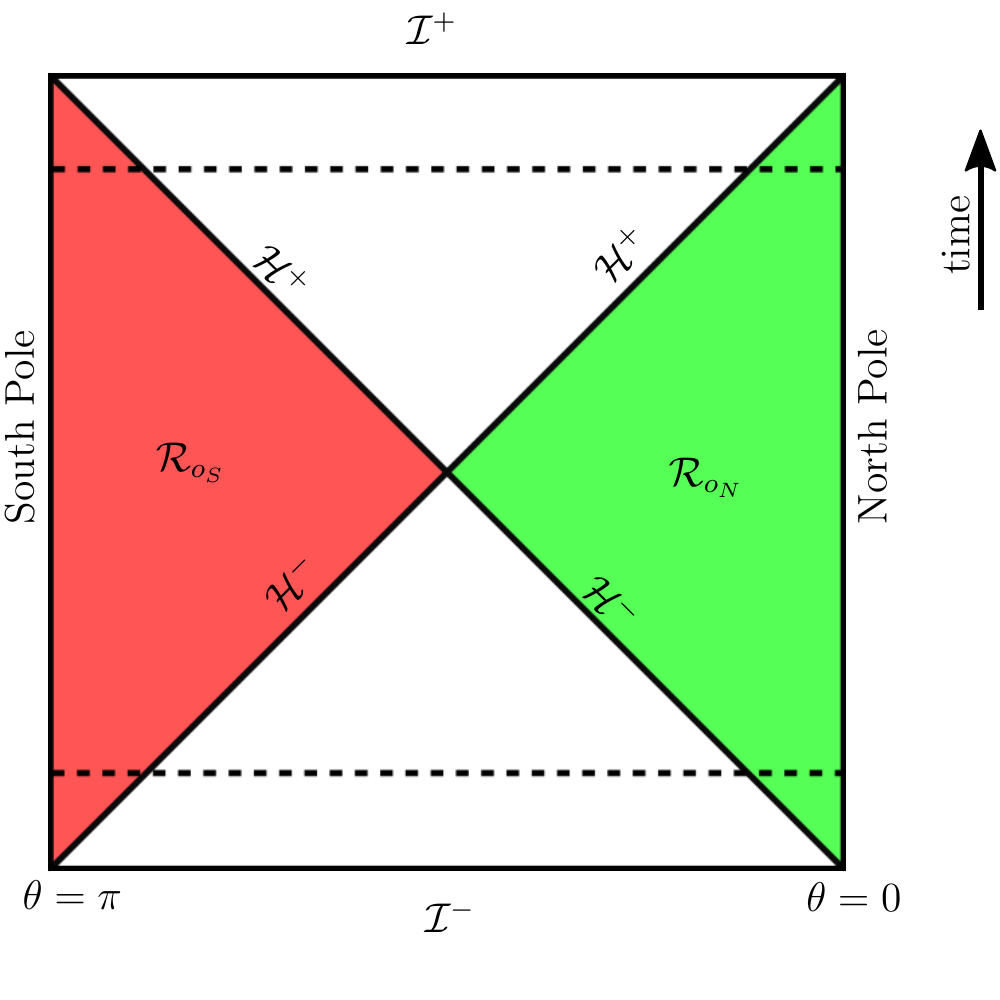}
	    \caption{The entangled Rindler-like wedges in $\deS$   corresponding to observers at the north and south pole. The dashed lines correspond to the boundaries of the slabs we consider.}
	    \label{fig:wedges}
          \end{figure}

In $\deS$, one wishes to calculate the SSEE between the two Rindler-like wedges which intersect at the
bifurcate horizon.   The $\deS$  metric is \cite{hawking_ellis_1973} 
 \begin{equation}
 ds^2= -d\tau^2+l^2\cosh^2(\tau/l)\,d\Omega_{d-1}^2, 
 \end{equation}
 where $-\infty<\tau<\infty$ and $l$ is the $\deS$  radius.  Using $\cosh(\tau/l)=\frac{1}{\cos T}$ it can alternatively be written as 
\begin{equation}
 ds^2=\frac{l^2}{\cos^2T}\left(-dT^2+d\Omega_{d-1}^2\right), 
 \label{confmetric}
\end{equation}
where $-\pi/2<T<\pi/2$, which is conformal to the round cylinder $S^{d-1}\times [-\pi/2,\pi/2]$. As shown in  the
conformal diagram in Figure \ref{fig:wedges}, associated with any time-like observer $o$ is a future/past horizon
$\cH_{\pm} = \partial (J^\pm(\gamma_o))$ where $\gamma_o$ is the world line of $o$. The Rindler-like wedge $\cR_o \equiv
J^+ (\gamma_o)\cap  J^- (\gamma_o))$ has a boundary which intersects $\cH_+$ and $\cH_-$ at a bifurcate horizon, whose
area is $A=4 \pi l^2$ in $4$d. 
Let us assume that the observer is at the south pole $o_S$. The Rindler-like wedge $\cR_{o_N}$ associated with its antipode at the
north pole, $o_N$, is then the  complement of $\cR_{o_S}$. The SSEE we wish to calculate is from the
entanglement between these two identical  Rindler-like wedges, which should therefore also satisfy complementarity. 

{The EE for $\deS_2$}  should have the same form as that for flat space. It contains a logarithmic UV cutoff dependence and in the case with two boundaries is given by \cite{solod, cardy}
\begin{equation}
  S= \frac{1}{3} \ln \biggl( \frac{l}{a} \biggr) + b,
  \label{2dentropy} 
\end{equation}
where $a$ is a  UV cutoff and $b$ a non-universal constant.  For $\deS_4$, the entropy-area relation, which is the same as the Bekenstein-Hawking  entropy, is expected to be \cite{gibbons, Bousso:2002fq}
\begin{equation} 
S = \frac {A}{\ell_p^2}=\frac{ c^3}{4 G\hbar} A = \pi l^2\indent \,\,    (\textrm{for}\,\, G=\hbar=c=1).
    \label{sh}
\end{equation} 
 It is with these formulae that we
will compare our results and ask if the causal set SSEE we find can account for the expected behaviour. It is understood that when \eqref{sh} is compared with the EE, the area of the entangled region in the EE is in units of the UV cutoff.

\subsection{Causal Set SSEE}
\label{ssee_in_cs}

Next, we set up the calculation of the horizon SSEE in a causal set approximated by  $\deS$. We refer
the reader to the literature on causal sets \cite{Bombelli:1987aa,Dowker:2005tz,Surya:2019ndm} and Appendix \ref{cs.app} for more 
details.


Given a finite volume $V$ region of a globally hyperbolic spacetime $(\mathcal{M},g)$, an ensemble of causal sets can be obtained
from it via a Poisson sprinkling at density $\rho$, where  the number of causal set elements $N$ is a random variable
whose average is given by 
\begin{equation}
 \langle N \rangle =\rho V. 
\label{nv}
\end{equation}
Causal sets obtained from a sprinkling are said to be continuum-like, and we will denote this by
$C \sim (\mathcal{M},g)$. For a particular realisation $C$, we denote by $C_O$ the sub-causal set approximated by the
subregion $O \subset \mathcal{M}$,  and its cardinality by $\Ns$.

Since our calculations are numerical, we are limited by the size $N$ and hence to finite volumes $V$ of $\deS$ .  As in \cite{Surya:2018byh} we pick a symmetric ``slab'' of $\deS$  with  $T \in [-h, h]$, so that in $4$d
\begin{equation}
  V_{slab}=\frac{4 \pi^2 l^4}{3} f(h), \quad f(h)=\tan h \biggl( \cos 2 h + 2 \biggr)\sec^2 h.
  \label{eq:slabvol}
\end{equation}
The causal sets we obtain from this sprinkling therefore have  finite $N$.

Before defining the causal set SSEE, it is useful to translate the continuum entropy-area relation to the discrete
  one,  by identifying the UV cutoff as the discreteness length $\rho^{-1/d}=\sqrt[d]{V/N}$. Replacing $\ell_p$ in \eqref{sh} with the causal set cutoff, we have in $d > 2$ 
\begin{equation}
 \langle \mathrm{S}^{(c)} \rangle= \rho^{\frac{2}{d}} \frac A4. 
  \label{eq: causalsetentropy}
\end{equation}
Using the slab volume \eqref{eq:slabvol} in  $\deS_4$,  this translates into the discrete entropy 
\begin{equation}
\mathrm{S}^{(c)} = \frac{1}{2} \sqrt{\frac{3}{f(h)}} \sqrt{N}.  
    \label{4dcsentropy}
\end{equation}

In $d=2$, the discrete entropy is given by taking the  cutoff  $a$ in \eqref{2dentropy} to be  $\sqrt{\frac{1}{\rho}}=\sqrt{\frac VN}$. In $\deS_2$, therefore, the discrete entropy should take the universal $d=2$ form  
\begin{equation}
    \mathrm{S}^{(c)}  = \frac{1}{6} \ln N +b .
    \label{2dcsentropy}
  \end{equation}

We now review the definition of the SSEE  associated with a Gaussian scalar field on a causal set $C$.  We begin with the discrete 
Pauli-Jordan function $i\Delta_C(x,x')$ which is  the difference between the causal set retarded and advanced Green functions
$G_{R,A}(x,x')$, for $x,x'\in C$, defined using the order relations in $C$.  The SJ prescription then associates a
unique state, or  Wightman function $W_C(x,x')$ as the positive part of $i\Delta_C(x,x')$ (see Appendix \ref{green.app}).
Next, consider any
causally convex subset $C_O \subset C$  and the  restrictions  $i\Delta_{C_O}(x,x'), W_{C_O}(x,x')$  of
$i\Delta_C(x,x')$ and $W_C(x,x')$ to $C_O$. Importantly, although $W_C(x,x')$ is a pure state, this is not true of
$W_{C_O}(x,x')$ which is not the positive part of $i\Delta_{C_O}(x,x')$. 
The simplest form\footnote{It is possible for example to have finite $N$ corrections to this formula, which vanish in
  the continuum limit.}  that the  causal set SSEE $\sseec$ takes is then 
\begin{equation}
  \sseec=\sum_{\mu} \mu \, \ln |\mu|,  \quad W_{C_O} \circ v=i\mu \Delta_{C_O}\circ v,\indent \Delta_{C_O} \circ v\neq 0,  
\label{s4c}
  \end{equation} 
where for $x \in C_O$,  $A_{C_O} \circ v (x) \equiv \sum_{x'\in C_0} A(x,x')v(x')$.   We will henceforth refer to the above equation and its continuum counterpart as the SSEE equation and $v$ and $\mu$ as {\sl generalised} eigenvectors and eigenvalues.  
Note  that in adapting the continuum formula to the causal set, we have retained the {strict} requirement that $v$ cannot lie in the kernel of $\Delta_{C_O}$.

For a causal set with no continuum counterpart, there is no unique or ``natural'' choice of $G_{R,A}(x,x')$,  and
subsequently no unique SJ vacuum and SSEE. Since an area law for the SSEE  is geometric, one expects such a relation to hold
only for causal sets which admit a geometric interpretation, i.e., a continuum approximation. Thus, although the SSEE  given by \eqref{s4c} can be
calculated for any causal set along with a choice of $G_{R,A}(x,x')$,  an area law makes sense only for continuum-like causal sets. By comparing  with the
continuum, the causal set Green functions $G_{R,A}(x,x')$ can be obtained in a class of continuum-like causal sets, including those approximated by $\deS$   \cite{Johnston2008,NomaanX2017}. This makes it possible to
explicitly calculate the causal set SSEE for the $\deS$  horizon.

An important aspect of the calculation of the EE is the introduction of a UV cutoff, which renders it finite. An unregulated quantum field in the continuum consists of infinitely many UV degrees of freedom which would yield an unbounded EE. When the regulated EE satisfies an area law, it is proportional to the spatial area of the entangled regions in units of the UV cutoff. This gives the scaling $S\propto a^{2-d}$ (for $d>2$), where $a$ is the UV cutoff in length dimensions and $d$ is the spacetime dimension.\footnote{For a heuristic argument for why the EE will in general be proportional to the spatial area of the entangling surface in units of the UV cutoff see \cite{Chandran:2015vuk}.} This is also true in the case of quantum theories with local interactions \cite{Eisert:2008ur}.

The causal set provides a natural cutoff length scale $a=\rho^{-1/d}\propto N^{-1/d}$. Based on this,  the
expected UV-dependence of the entanglement entropy $S$ of a scalar field in various dimensions  is as shown in the table
below, with \eqref{4dcsentropy} and \eqref{2dcsentropy} being special cases.  Since the leading area term in $d=2$ is a constant (as the spatial boundary of the entangling region is one or
two points), one also considers the subleading contribution $c_1\ln a$, where $c_1$ is a universal constant.\footnote{When
  the spatial boundary is a single point $c_1=-1/6$, and when it is two points  $c_1=-1/3$.} 
 \begin{table}[h]
\label{comtable}
\begin{centering}
\begin{tabular}{|c|c|c|}
\hline
 {\text{Spacetime Dimension}} & $S(a)$ &  $S(N)$\\ \hline
 		$d=2$ &  $c_1\ln a+\text{const}$ &  $-c_1\ln \sqrt{N}+\text{const}$\\ \hline
 		$d=3$ &  $1/a$ &  $N^{1/3}$ \\ \hline
 		$d=4$ &  $1/a^2$ &  $\sqrt{N}$ \\
 \hline
\end{tabular}
\caption{The dependence of the entropy $S$ on the UV length cutoff $a$ and causal set size $N$.}
\end{centering}
\end{table}

If instead  $S(N)\sim N$,  this means 
that $S$ satisfies a  spacetime {\it volume} rather than an area law. Interestingly, we will see that this is what commonly happens in the causal set when we compute the SSEE without any truncations. 

\subsection{Review of Causal Set SSEE for Nested Causal  Diamonds in $\mink^2$}
\label{2dmink_review}

In order to set the stage we review the results of  \cite{Sorkin:2016pbz} for causal sets approximated by the nested
causal diamonds $\diam_\ell^2\subset \diam_L^2\subset \mathbb{M}^2$, with side lengths $2L>2\ell$ (see Figures \ref{diamonds}
and \ref{fig:2ddiamonds}).   In this special case, one can make comparisons with the continuum results of
\cite{Saravani:2013nwa} which made use of the fact that the continuum SJ modes for $\diam_L^2$ are explicitly known \cite{Johnston:2010su}:
	\begin{eqnarray}
	f_k(u,v)=e^{iku} - e^{ikv} &|& k=\frac{n\pi}{L},\quad\, n\in\mathbb{Z^\pm}\nonumber\\
	g_k(u,v)=e^{iku} + e^{ikv}-2\cos kL &|& k\in \text{ker} (\tan(kL)-2kL)\nonumber\\ && \xrightarrow[]{m\rightarrow\infty}\bigg(m-\frac{1}{2}\bigg)\frac{\pi}{L}\approx\frac{m\pi}{L}\,,\,\,m\in\mathbb{Z^\pm}. \nonumber 
	\end{eqnarray}
In the UV limit, i.e.,  for large $k$,  the SJ spectrum takes the simple form $\lambda_k=\frac{L}{k}$ for both sets of
modes. {In this limit, these modes} moreover become linear combinations of the same plane waves, but are out of phase. Thus the UV part of the SJ spectrum for both modes can be characterised by an integer $n$, with $k=\frac{n \pi}{L}$.    

For a causal set approximated by $\diam_L^2$, the SJ spectrum was calculated using the $d=2$ causal set retarded Green function (see Appendix \ref{green.app}) \cite{Sorkin:2016pbz}. Figure \ref{fig:knee} shows a comparison of the continuum and causal set SJ spectra for $\diam_L^2$ which match up to the characteristic ``knee'' mentioned in the Introduction. As the sprinkling density $\rho=N/V$ increases, the knee in the causal set SJ spectrum occurs at larger $k$ values.   

The continuum SSEE was calculated in \cite{Saravani:2013nwa} for $\diam^2_\ell \subset  \diam^2_L$ using a cutoff
$a=1/k_{\mx}$ and shown to satisfy the expected ``area'' law of \eqref{2dentropy}. 
However, the analogous  calculation in the causal set, yielded a volume law,  $\sseec \propto N$,  rather than an area law \cite{Sorkin:2016pbz}. 
        
This surprising feature, which is markedly different from the continuum result, can be traced to the shape of the causal
set SJ spectrum. As evident in Figure \ref{fig:knee}, beyond the knee the causal set SJ spectrum contains a large number
of near zero eigenvalues, which are absent in the continuum. In the causal set SSEE,  \eqref{s4c}, the generalised
eigenvector $v$ is required to lie outside the kernel of $\Delta_{C_O}$. However, because of the nature of the causal
set discretisation, fluctuations near the cutoff scale $\rho^{-1}$ can yield  eigenvectors that are ``almost'' but not
strictly in the kernel. This is true in general of the  discrete-continuum correspondence: as one  gets closer to the
spacetime discreteness scale $\rho^{-1}$, the relative fluctuations get larger. Thus, it is reasonable to expect that at
such scales, the causal set SJ spectrum will deviate significantly from the continuum. 

Indeed, as  shown in \cite{Sorkin:2016pbz}, { truncating}   the SJ spectrum of both $i\Delta_C$
as well as $i\Delta_{C_O}$ around {this}  knee has the effect of giving back the expected $d=2$ ``area law'' as in the continuum. Such a truncation can be motivated by appealing to the fact that the SJ modes $f_k,g_k$ are combinations of plane wave modes with wavenumber $k=\frac{2 \pi}{\nu}$, where $\nu$ is the wavelength. The causal set discreteness then gives a natural choice for the minimum wavelength $\nu_\mn \sim \rho^{-1/2}=2L/\sqrt{N}$. 
Since $k\sim \frac{ n \pi}{L}$ for large $k$, this suggests a truncation to retain as many modes as  $n_\mx \sim
\sqrt{N}$. The dimensionless causal set  {SJ eigenvalue} $\lambda^{cs}$ is related to the dimensionful continuum
{SJ eigenvalue}  $\lambda$ by $
\lambda^{cs}=\rho^{\frac{2}{d}}\lambda$. This means that $\lambda_\mn=\frac{L^2}{ \pi n_{max}}$ corresponds to $\lambda^{cs}_\mn \sim \frac{\sqrt{N}}{4 \pi}$ when we  
truncate the SJ spectrum with   $n_{\mx} \sim  \sqrt{N}$. 
     
The choice of $\sqrt{N}$ modes can also be justified by appealing to another aspect of the continuum picture. In the conventional  spatial way of understanding a quantum field and its EE, the field is quantised on a spatial Cauchy hypersurface and the contributions to the EE come from the field modes on that Cauchy hypersurface. In the continuum we do not expect to have {\it more} field modes contribute to the SSEE than in the spatial case, when we are working with domains of dependence. While the space of our solutions $\dim(\Delta)=N$ is larger, the space of independent solutions given by the $\mathrm{Im}(i \Delta)$ should remain the same as in the spatial picture. We expect the latter to be given in terms of the spatial volume (here the length) of the Cauchy hypersurface, so that the number of non-redundant solutions $\sim \sqrt{N}$ (where we have singled out the time-symmetric $t=0$ diameter of the causal diamond). Alternatively, since $\lambda$  has a dimension of $(\mathrm{length})^2$, we may assume that it is more generally the product of an IR scale and a UV scale,  $\lambda^{cs}_\mn\sim \rho^{\frac{1}{2}}L\sim \sqrt{N}$.  We  note  that since the number of the eigenvalues is $\sim N$, the reduction to $\sqrt{N}$ modes is a very non-trivial restriction.   

Thus, we have a {\sl number truncation} characterised by $n_{\mx}$ which gives the number of (largest in magnitude) eigenvalues that are retained, or alternatively, a {\sl magnitude truncation} $\lambda^{cs}_\mn$ which gives the minimum magnitude of the eigenvalues that are retained. These are related in $\diam_L^2$ by 
\begin{equation} 
  \lambda_\mn^{cs} = \frac{N}{4\pi n_{\mx}},
  \label{eq:magtrunc}
          \end{equation} 
but this relation may not hold more generally.

Once the truncation scheme is decided, the truncation needs to be implemented {\it twice}. This is the {\sl
  double truncation} followed in \cite{Sorkin:2016pbz}  which we describe in some detail below for the specific case
of $\diam_\ell^2 \subset \diam_L^2$. {Our notation is a little heavy for the sake of clarity, but we will shed it
  for simpler notation subsequently.} 

The first truncation $n_\mx\sim\sqrt{N}$ or $\lambda_\mn^{cs} \sim \frac{\sqrt{N}}{4\pi}$  is on the SJ  spectrum in
$\diam_L^2$, which therefore also truncates the {operator $i\Delta_L^t$ and therefore the} SJ  Wightman function
$W_{L}$ to $W_L^t$.   After the first truncation the region beyond the knee in the SJ spectrum of $i\Delta_L$ 
{is removed, leaving behind a residual} power law {behaviour}. {Next,} when $i\Delta_L^t$ is restricted to
$\diam_\ell^2$, i.e., $i\Delta_\ell^t(x,x')\equiv i \Delta_L^t(x,x')|_\ell$ {the}  knee reappears once again in the
spectrum of the corresponding integral operator {$i\Delta_\ell^t$ in $\diam_\ell^2$}. Hence a second
truncation with $n^\ell_\mx\sim\sqrt{N_\ell}$ is necessary {in the spectrum of}  $i \Delta_\ell^t$, {which we
  denote by  $i \Delta_\ell^{tt}$}.  Finally, the
restriction $W_l^t\equiv W_L^t|_\ell$  of $W_L^t$ to $\diam_\ell^2$,  must then be further projected onto this smaller
(double) truncated subspace of the eigenbasis of $i \Delta_\ell^t${, to give us $W_l^{tt}$}. {Note that $i
  \Delta_\ell^{tt}$ is {\it not} the operator obtained after truncating the spectrum of the
  Pauli-Jordon operator $i
  \Delta_\ell$ in $\diam_\ell^2$.} 
    
The reappearance of the knee in the spectrum of $\mathrm{Im} (i \Delta_\ell^t)$ can be traced to the fact that the Pauli-Jordan integral {\it operators} $i\hat{\Delta}_L$ and $i\hat{\Delta}_\ell$ are defined over different integral domains and hence the  spectrum of  $i\hat{\Delta}_\ell$ cannot be obtained from a restriction of that of $i\hat{\Delta}_L$. This ``non-locality'' is an important feature of the SJ vacuum. 
{Most importantly}, without this second truncation, the full set of ``near zero'' elements in $\mathrm{Im} (i \Delta_\ell^t)$ is not removed and this gives rise to a too-large SSEE.  

This gives us a template for implementing the double truncation procedure more generally, for any $C_O \subset C$. Thus, the first truncation is performed on  the SJ spectrum of $i\hat{\Delta}_C$ to give the  truncated operator  $i\hat{\Delta}^t_C$, and its associated Wightman function $W_C^t(x,x')$. The restriction of the truncated Pauli-Jordan function $i\Delta^{t}_{C_O}(x,x')= i{\Delta}^t_C(x,x')|_{C_O}$ corresponds to an  operator $i\hat{\Delta}^{t}_{C_O}$ in $C_O$, i.e., for $x \in C_O$,  $i\hat{\Delta}^{t}_{C_O}\circ v(x) = i \sum_{x'\in C_O}\Delta^{t}_{C_O}(x,x')v(x')$. 
    
{The second truncation is then performed on the spectrum of $i\hat{\Delta}^{t}_{C_O}$, which yields the operator
$i\hat{\Delta}_{C_O}^{tt}$, as well as the  projection  $W_{C_O}^{tt}$ of the restriction $W_C^t(x,x')|_{C_O}$ to this
second truncated eigenbasis. Thus} the double truncated SSEE {version}  of \eqref{s4c} is 
\begin{equation} 
\sseec=\sum_{\mu} \mu \, \ln |\mu|,  \quad W_{C_O}^{tt} \circ v=i\mu \Delta_{C_O}^{tt}\circ v,\indent
\Delta_{C_O}^{tt} \circ v\neq 0,  \label{s4ct}
\end{equation}
where $tt$ denotes the double truncation procedure described above. {We now  drop the ``${tt}$''
  superscript for simplicity of notation, and refer to the spectrum as either truncated or untruncated.}  

\subsection{Generalised Truncation Schemes}
\label{gen_trunc}
	
In what follows, we discuss ways in which to generalise the truncation procedure in $\diam_L^2$ without explicit
knowledge of the SJ spectrum in the continuum. Out of the several possibilities, the ones that would closely mimic the
continuum would be those that satisfy  an area law relation for the SSEE compatible with the Bekenstein-Hawking entropy, as well as complementarity.

We consider causal sets obtained by sprinkling into the finite volume ``slab''  between $[-h, h]$ in $\deS$. As discussed in Section \ref{areas}, the south and north Rindler-like wedges $\cR_{o_S}$ and $\cR_{o_N}$ are complementary to each other, and intersect only at the equator of the $t=0$ $3$-sphere. In the $\deS$  slab, these regions have hyper-hexagonal boundaries (see Figure \ref{fig:wedges}). Both $\cR_{o_S}$ and $\cR_{o_N}$ are also time-symmetric and are the domains of dependence of time-symmetric $t=0$ Cauchy slices, which are the Southern and Northern hemispheres of the 3-sphere, respectively.

Since the SJ spectrum of the hyper-hexagon is not known, we cannot resort to comparisons with the continuum as in the
nested diamonds. {In the course of our investigations we tried a very large number of different truncation schemes. Of these
  we} focus on two particular schemes which we think are physically motivated and simple to generalise and
  moreover, give an SSEE which satisfies an area law.  

The first choice we make is an estimation of the number truncation $n_{max}$, inspired by the {nested $d=2$
  diamonds, where}   $n_{\mx}=\sqrt{N}$ for each of the two sets of modes.  { This was motivated by the fact that the 
number of modes should be proportional to the spatial volume of a Cauchy hypersurface.} A natural generalisation of this is
\begin{equation}
  n_{\mx} = \mathrm{\con} {N}^{\frac{d-1}{d}}.
  \label{eq:numtrunc} 
\end{equation}  
Note that {the identification of the spatial volume} is neither uniquely nor
covariantly defined, and hence  there is no unique {choice of $\con$}; in particular one can deform the Cauchy hypersurface to one that has arbitrarily small spatial volume.  
{ In our investigations of the nested causal diamonds in $\mink^4$ (Section \ref{mink.app}) we experimented with several values of $\con$, including that corresponding to the volume of the time symmetric
  slice. In the de Sitter case, this latter factor turns out to be too large, leading to too small a truncation. As a result,
  we focus here only on values of $\con$ which give the most reasonable results,  i.e., $\con=1,2$.}

\begin{figure}[!h]
	\centering
	\includegraphics[width=0.65\textwidth]{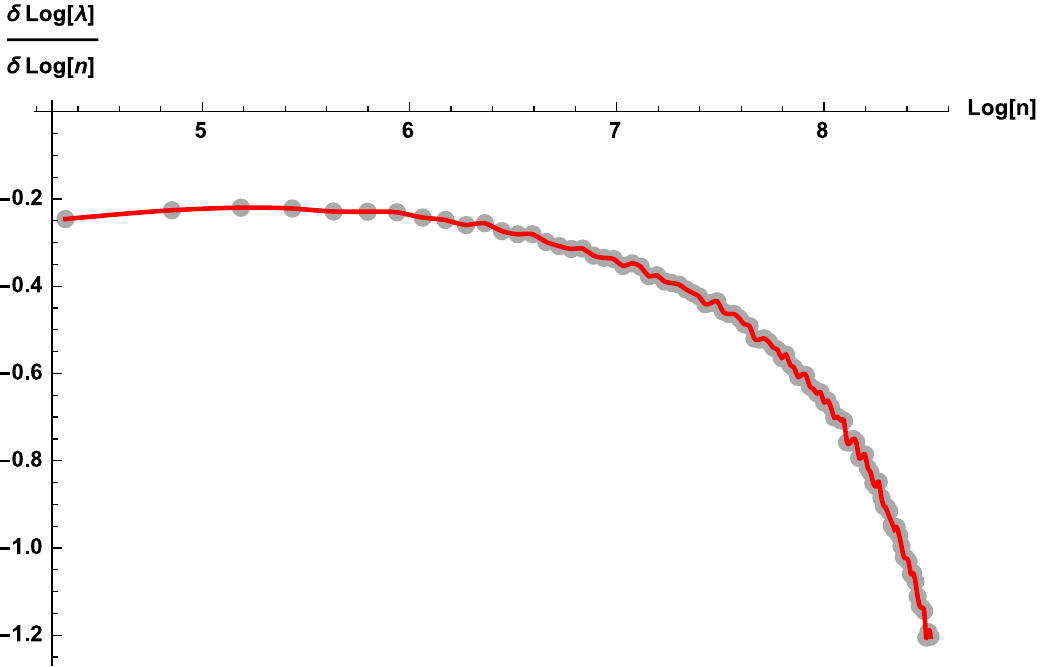}
	\caption{Slopes of the log-log SJ spectrum in $\deS_4$. Data points are binned averages and the curve is an interpolating function fit to the data.}
	\label{fig:slopes}
\end{figure}

Our second choice is a new truncation scheme, which we dub the linear  scheme.  Since the SJ spectrum in the causal
set is a power law and therefore linear in the log-log plot, up to a characteristic knee,  it is reasonable to truncate the spectrum at the
point where this linear regime ends.  This requires an estimation of  the end of the linear regime in the log-log
plot. One method
is to use the change in the slope of the logarithms of the
data. We implement this in the following way: First, the logarithms of each $n^{th}$ eigenvalue are taken along with the logarithm of its label $n$. Then the
slope of the line between each nearest neighbor pair\footnote{An  alternative method, which yields similar results, is
  to take the slopes of more than than a pair (say, every 50) of nearest eigenvalues.} of data points is computed. Due
to fluctuations in the causal set data, these slopes also fluctuate when going from one pair's slope to the next, even
in the (approximate) power law regime. In order to smooth out these fluctuations, the slopes are binned and
averaged. Then, a smooth interpolating function is fit to the averaged slopes as shown in Figure \ref{fig:slopes}. This interpolating function can then be
used to track the drop in the slope and set the truncation number or magnitude. {The region of nearly constant
  (negative) slope $m$ is first  identified, and the estimation of the knee corresponds to a drop to a more negative
  $m'$. A choice is then made of the fractional drop $\delta=\frac{m-m'}{m}$ to obtain the knee.} 
We take the magnitude of the eigenvalue {at this estimated knee} as our magnitude
truncation, or the number $n_{\mx}$ (rounded to the nearest integer) at which this happens as our number
truncation. {We have explored various choices of
  $\delta$, also allowing it to be different in the slab and in the Rindler-like wedge. } 

An advantage of the linear truncation over the {generalised number}   truncation \eqref{eq:numtrunc} is
that it is covariantly defined, without appealling to any features of a Cauchy surface and {the associated}
ambiguity of {choosing} a proportionality constant. There is of course the fine tuning that comes with the
  choice of $\delta$ and the hope is to be able to find a suitable range of values, 
as much as the quality of data allows.

In all the cases we study, the numerically generated causal set SJ spectrum can additionally be used to estimate the power law behaviour of $\lambda^{cs}$ as a function of $n$. Rescaling the spectrum by $\rho^{-\frac{2}{d}}$ collapses the data in the linear regime, so that 
\begin{equation}
   \rho^{-2/d}\lambda^{cs}=\frac{b}{n^a},
  \label{eq:csmagtrunc}
\end{equation}
where the exponent $a$ and the constant $b$ can be determined empirically. For $\diam_L^2$, for example, $a=1$ and
$b=1/(4 \pi)$. For the $\deS_2, \deS_4 $ slabs and associated Rindler-like wedges 
these values are given in the following table{, where the slab height has been chosen to be $h=1.2$.}

\begin{table}[h]
\begin{centering}
\begin{tabular}{|c|c|c|}
\hline
 {Spacetime} & Slab &  Wedge\\ \hline
 		$\deS_2$ &  $a\sim 1,\,b\sim 1.68$ &  $a\sim 1,\,b\sim 0.26$\\ \hline
 		$\deS_4$ &  $a\sim 0.25,\,b\sim 2.16$ &  $a\sim 0.36,\,b\sim 0.78$ \\ \hline
\end{tabular}
\caption{The values of parameters $a$ and $b$ in \eqref{eq:csmagtrunc} determined from the spectrum of $i\Delta$ in the regions considered.}
\end{centering}
\end{table}
This also allows us to translate $n_\mx$ (picked either by the number or linear truncation method) into a magnitude
truncation $\lambda^{cs}_\mn$ for this choice of $h$. We have not however studied the effect of varying $h$ on the parameters $a$ and $b$ and whether or not the spectrum can
  be collapsed to a universal form.  


\section{Results}
\label{results}   

The simulations presented here were performed using Mathematica on an HP Z-8 workstation with 320GB pooled RAM. For
larger $N$ values, {a significant fraction} 
of this pooled memory was used in {the simulation, when all the trials for fixed $N$ are parallelised}. The results
presented here are the culmination of {extensive exploration of various truncation schemes, including certain
  magnitude 
  truncations not described in Section \ref{gen_trunc}. Here we only present results from the two described in  Section
  \ref{gen_trunc} and for  choices of $\con$ and $\delta$ which best satisfy the 
  criterion of an area law compatible with the Bekenstein-Hawking entropy. As mentioned in Section \ref{gen_trunc}, the
  two Rindler-like wedges are indentical and hence complementarity should be automatically satisfied. In our investigations, we also calculated  the SSEE for a causal diamond in
  the slab spacetime, whose complement is not necessarily a causal diamond, but for this work we present results only  from  the Rindler-like wedges, {since
  these are of most interest for the $\deS$ horizons.}

\subsection{$\deS_2$}
\label{ds2}

In $\deS_2$,  the two complementary regions $\cR_{o_S}$ and $\cR_{o_N}$  are each conformal to causal diamonds.  The
simulation results we present are for a slab of $\deS_2$ of height $h=1.2$ into which we sprinkle  causal sets with
sizes $\langle N \rangle$ ranging from $2000$ to $16000$.

Figure \ref{fig:ds2notrunc} shows the
dependence of the SSEE with $N$ without truncating the SJ spectrum.  The SSEE clearly scales
linearly with $N$ and therefore obeys a spacetime volume law, as in the case of the $d=2$ nested diamonds \cite{Sorkin:2016pbz}. 
 \begin{figure}[!htp]
 	\centering
   \includegraphics[width=0.6\textwidth]{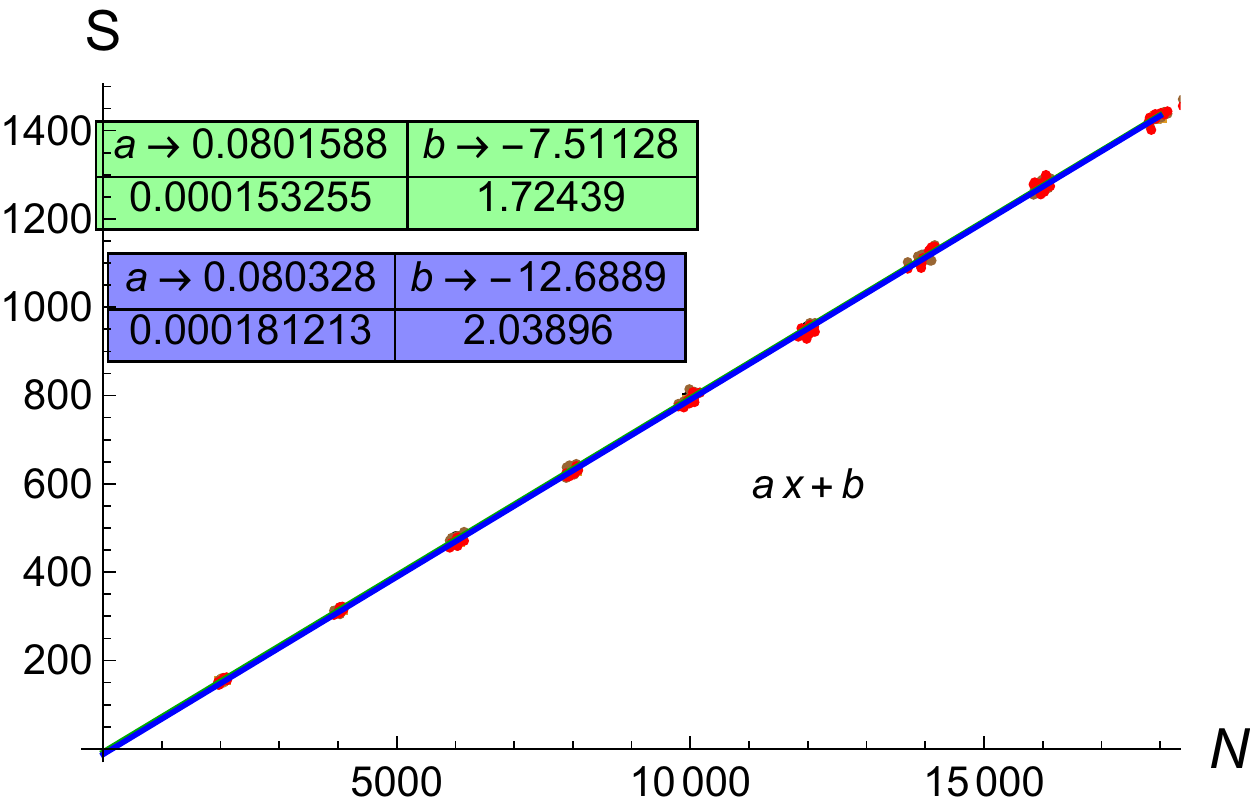}
  \caption{Untruncated SSEE vs. $N$ in the $\deS_2$ slab of height $h=1.2$ for the two
    Rindler-like wedges $\cR_{o_S}$ and $\cR_{o_N}$ (shown in green and blue).  The best fits are shown.}
   \label{fig:ds2notrunc}
  \end{figure}

 Next we implement the truncation schemes discussed in Section \ref{gen_trunc} for the SJ spectrum for a causal  region
 of cardinality $N_s$. For each  $\langle N \rangle$, we run  $10$ simulations for the number truncation while we run
$5$ simulations for the linear truncation. For the  latter,  the estimation of the linear regime is done for the SJ  spectrum
in the slab as well as for the SJ spectrum in the Rindler-like wedges.

 {For the}  number truncation \eqref{eq:numtrunc} we  work with {$\con$} values of $1$ and $2$, the latter being the analogue of the $2$d causal diamond truncation.\footnote{Note that while $n_\mx$ in our review of the $2$d causal diamond denoted the maximum number of modes of each family of $f$ and $g$ eigenfunctions, here we refer to it as the total number of eigenfunctions irrespective of  degeneracies. Hence the two-fold degeneracy of the $2$d diamond amounts to keeping a total of $2\sqrt{N}$ eigenvalues in the terminology henceforth.}
For the linear truncation scheme, different values of $\delta$ were explored. We found that the
   one most compatible with the area law is $\delta \sim 0.1$} 
 in both the slab and the Rindler-like wedge.

\begin{figure}[!htp]
  \begin{subfigure}[b]{0.5\textwidth}
  \includegraphics[width=\textwidth]{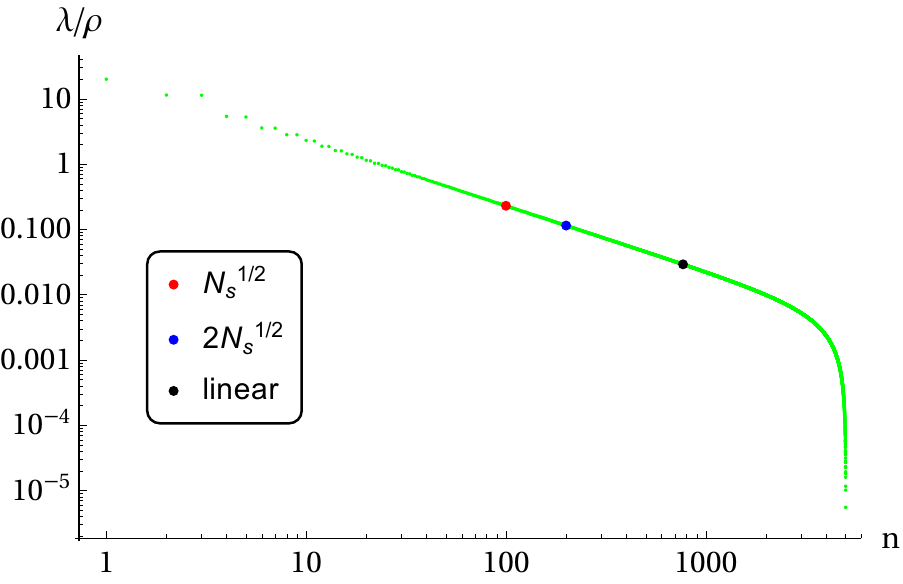}
  \caption{}
  \label{ds2spectrummarked}
  \end{subfigure}
  \begin{subfigure}[b]{0.5\textwidth}
  \includegraphics[width=\textwidth]{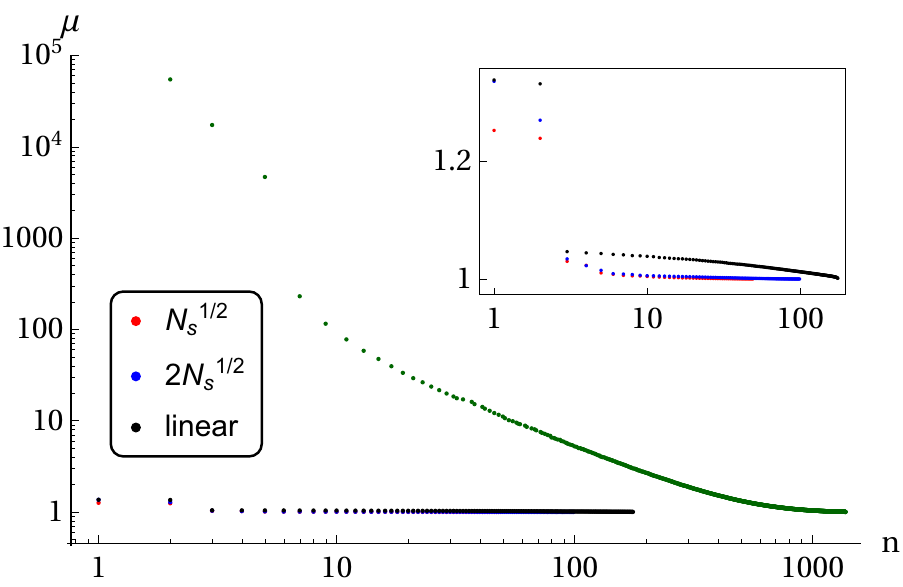}
  \caption{}
  \label{ds2 generalized spectrum}
  \end{subfigure}
  \caption{ (a) The SJ spectrum for an $N=10^4$ causal set sprinkled into the $\deS_2$ slab. Three
    different truncations choices are marked. 
    (b) The spectrum for the SSEE \eqref{s4c} with and without these truncations.}
  \end{figure}

 In Figure \ref{ds2spectrummarked} we show the log-log plot of the untruncated causal
 set SJ spectrum of the $\deS_2$ slab,  with these three choices for truncation marked. All three clearly lie in the
 linear regime, with the linear truncation being the closest to the knee. In Figure \ref{ds2 generalized spectrum} we show the
log-log spectrum of the generalised eigenvalue equation before and after truncation. What is striking is the drastic
reduction in not only the number but also the magnitude of the eigenvalues.  It is this feature that seems to make it
possible to recover an area law after truncation. 

\begin{figure}[!h]
  \begin{subfigure}[b]{\textwidth}
  \includegraphics[width=0.5\textwidth]{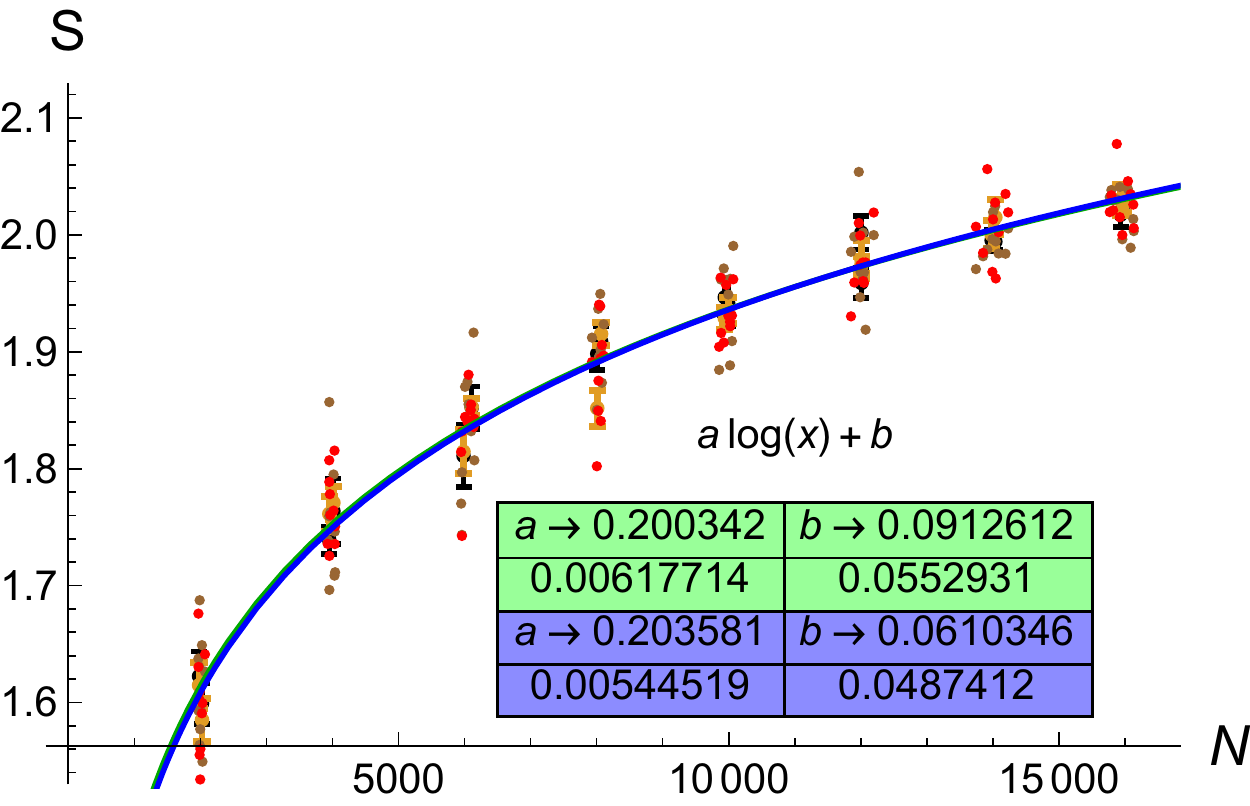}
  \includegraphics[width=0.5\textwidth]{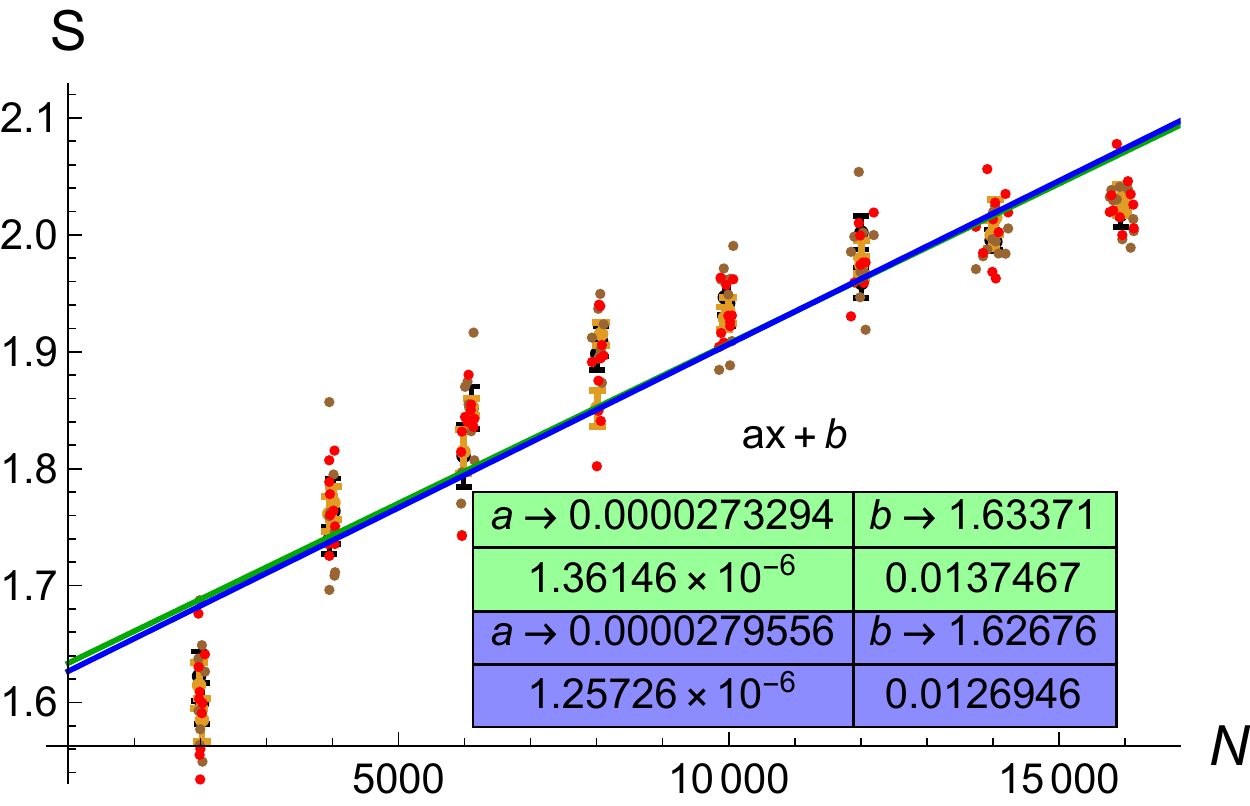}
  \caption{Number truncation with $n_\mx=N_s^{1/2}$}
  \label{ds2num}
  \end{subfigure}
\begin{subfigure}[b]{\textwidth}
	\includegraphics[width=0.5\textwidth]{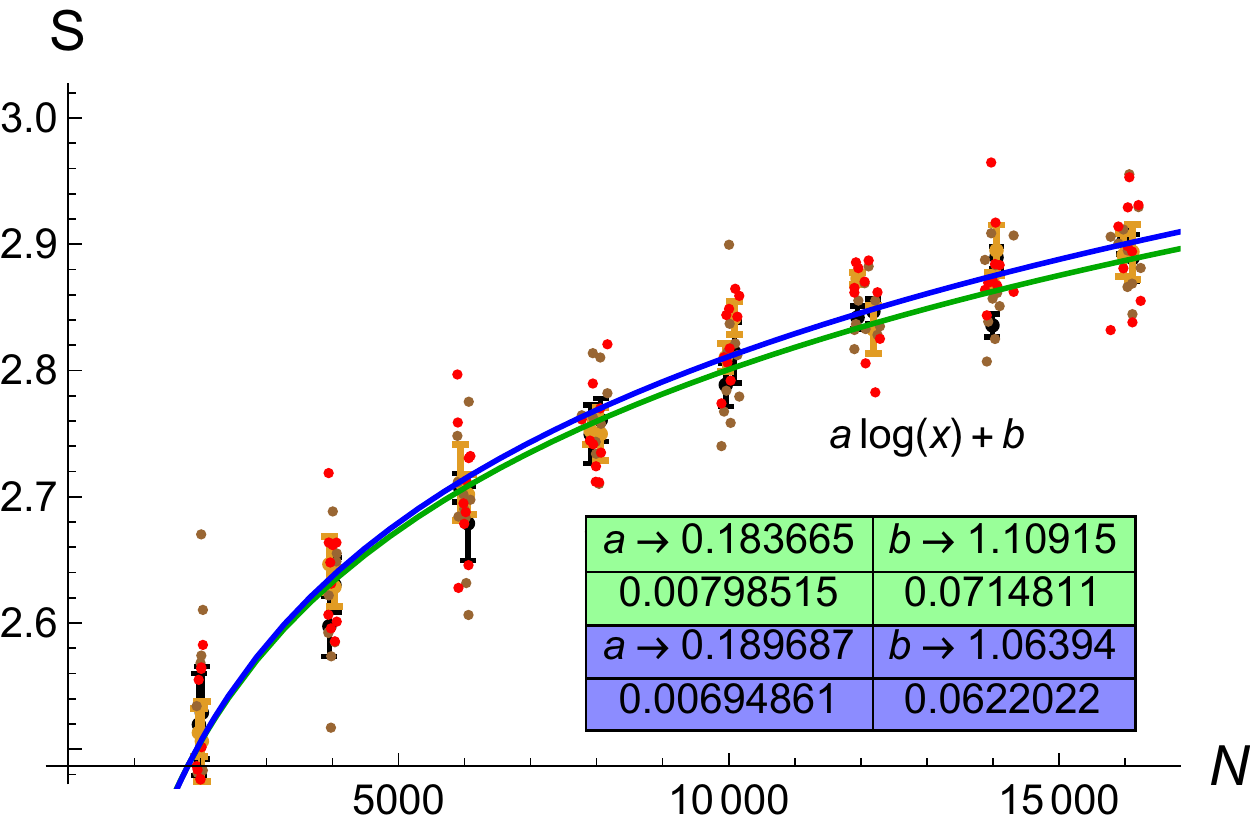}
	\includegraphics[width=0.5\textwidth]{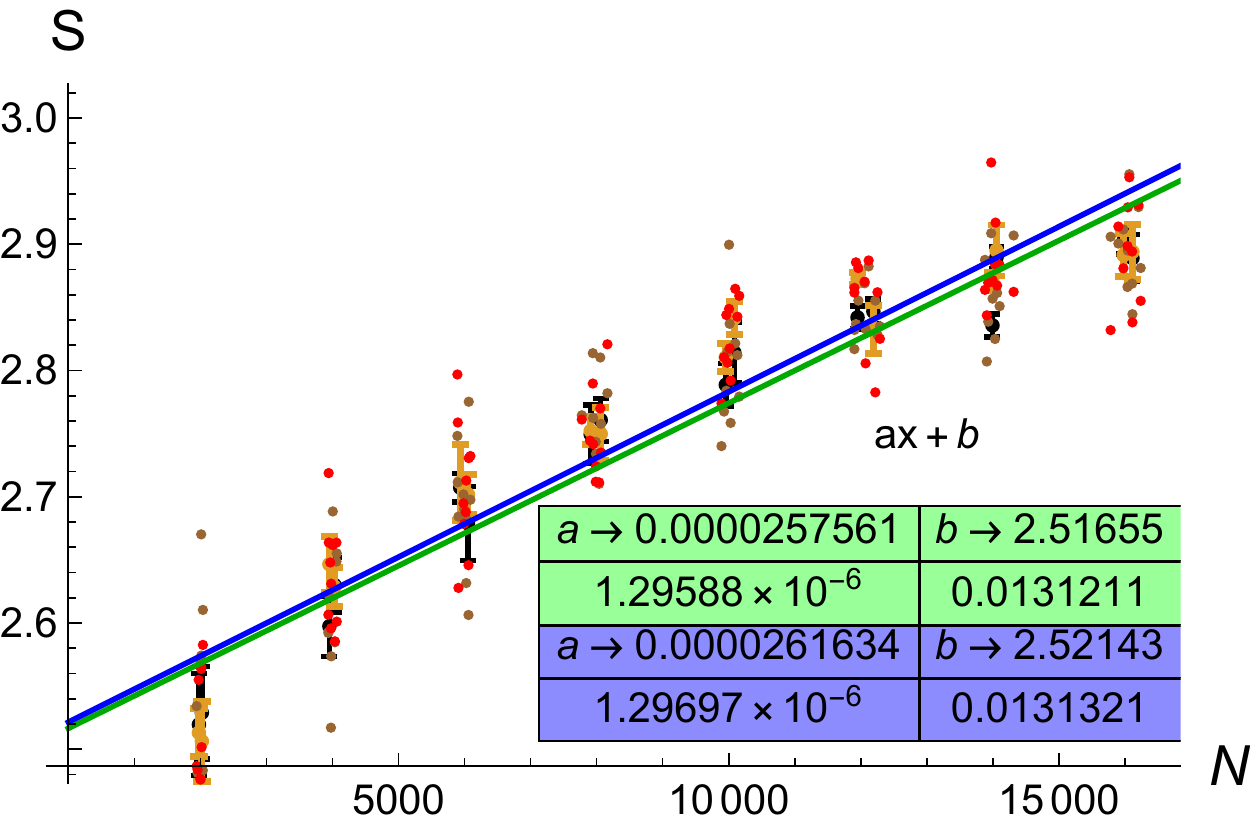}
	\caption{Number truncation with $n_\mx=2N_s^{1/2}$}
	\label{ds2fac2}
\end{subfigure}
  \begin{subfigure}[b]{\textwidth}
  \includegraphics[width=0.5\textwidth]{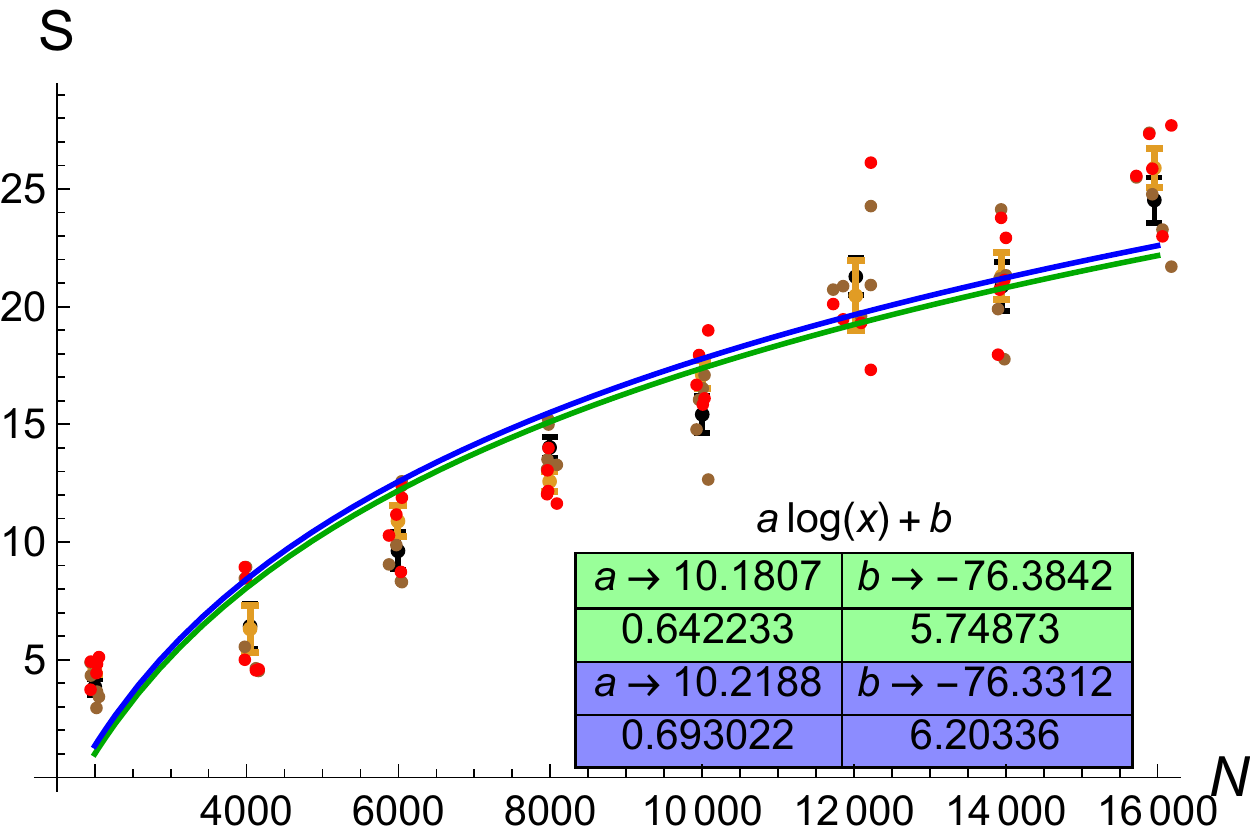}
  \includegraphics[width=0.5\textwidth]{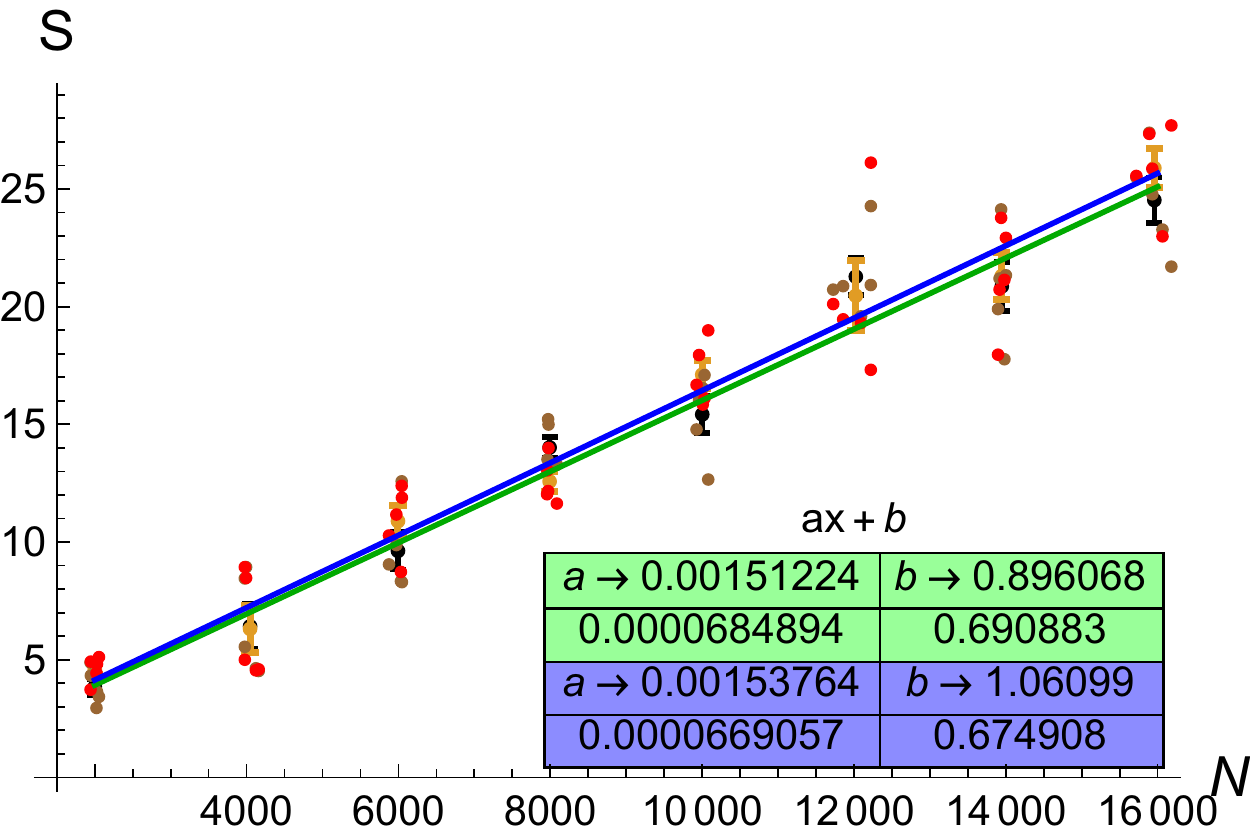}
  \caption{Linear truncation}
  \label{ds2linear}
  \end{subfigure}
  \caption{SSEE vs. $N$ with three different choices of truncation in $\deS_2$.  The green and blue represent the data for
    the two Rindler-like wedges. A comparison of the two fits $a\ln{x}+b$ and $ax+b$ is shown on the left and the right
    for each choice of truncation.}
  \label{ds2 entropy trunc}
\end{figure}

  Finally, in Figure \ref{ds2 entropy trunc} we show the SSEE calculated using the above three truncations for both
  $\cR_{o_S}$ and $\cR_{o_N}$.    For each truncation, on the left we show the fit to the logarithmic behaviour 
  \begin{equation}
    \sseec = a \ln N+ b,  
  \end{equation}
  (where the expected value of $a$ is  $1/6$) and on the right,  the fit to the volume behaviour $a N+ b$. {The
    errors in the best fit parameters are given below these values.}
  The fit and corresponding uncertainities are found using the least square method.
  We see in all three cases that the data has a high degree of scatter, which is also the case for the $d=2$ nested diamonds
\cite{Sorkin:2016pbz}  and seems to be a characteristic of $d=2$.  All cases are reasonably consistent with an area law,
but the linear truncation is  surprisingly more consistent with a volume law. All cases 
also satisfy complementarity up to Poisson fluctuations. 

From these results we conclude that the truncation that is closest to the expected EE values is the
choice $n_\mx=2\sqrt{N}$, with $a$ and $b$ values given in Figure \ref{ds2fac2}. This case gives  $a\sim0.18$ which is
closest to the expected value of $1/6$. 
 
\subsection{$\deS_4$}
 \label{ds4}
 
 The $\deS_4$ slab is again taken to have height $h=1.2$.  We consider causal set sprinklings with $\langle N \rangle$
 ranging from $2000$ to $16000$.
 
\begin{figure}[!h]
 	\centering
   \includegraphics[width=0.6\textwidth]{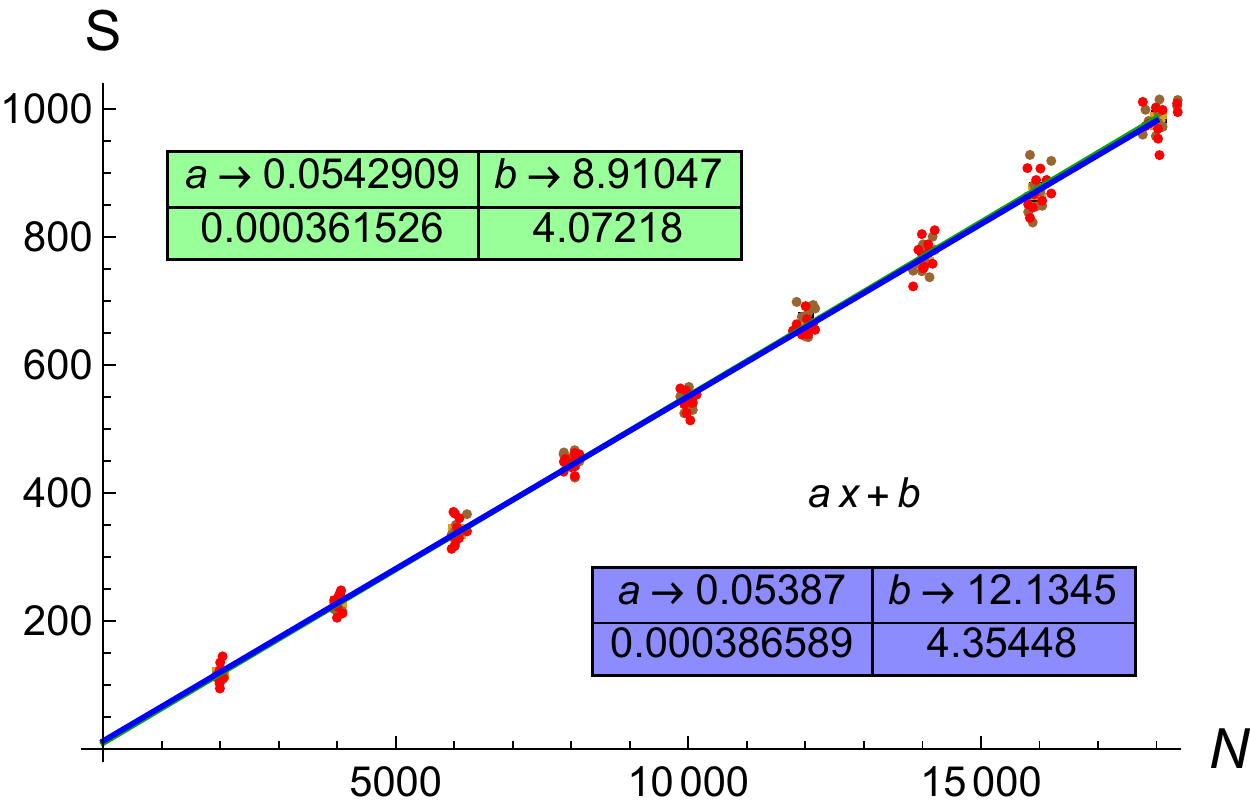}
  \caption{Untruncated SSEE vs. $N$ in a $\deS_4$ slab of height $h=1.2$ for the two
    Rindler-like wedges $\cR_{o_S}$ and $\cR_{o_N}$ (shown in green and blue).  The best fits are shown.}
   \label{ds4notrunc}
 \end{figure}

 In Figure \ref{ds4notrunc} we show the untruncated SSEE which again clearly scales linearly with $N$ and
 therefore obeys a spacetime volume law.

 We present results for three choices of truncations, the number truncations $n_\mx=N^{\frac{3}{4}},\,2 N^{\frac{3}{4}}$ and
 the linear truncation.  We run $10$ simulations for each fixed $\langle N \rangle$ for both types of truncation. 
As in $\deS_2$,  for the  latter,  the estimation of the linear regime is done for the SJ  spectrum
in the slab as well as for the SJ spectrum in the Rindler-like wedges. We {find that a choice of $\delta \simeq 0.15$
  for both the slab and the Rindler-like wedge spectrum gives the
  best results.}   
Note that since the knee is fairly sharp in the  log-log spectrum,  even a seemingly large tolerance does not lead to  very drastic
changes in the spectrum but does change the SSEE so obtained. 

\begin{figure}[!h]
  \begin{subfigure}[b]{0.5\textwidth}
  \includegraphics[width=\textwidth]{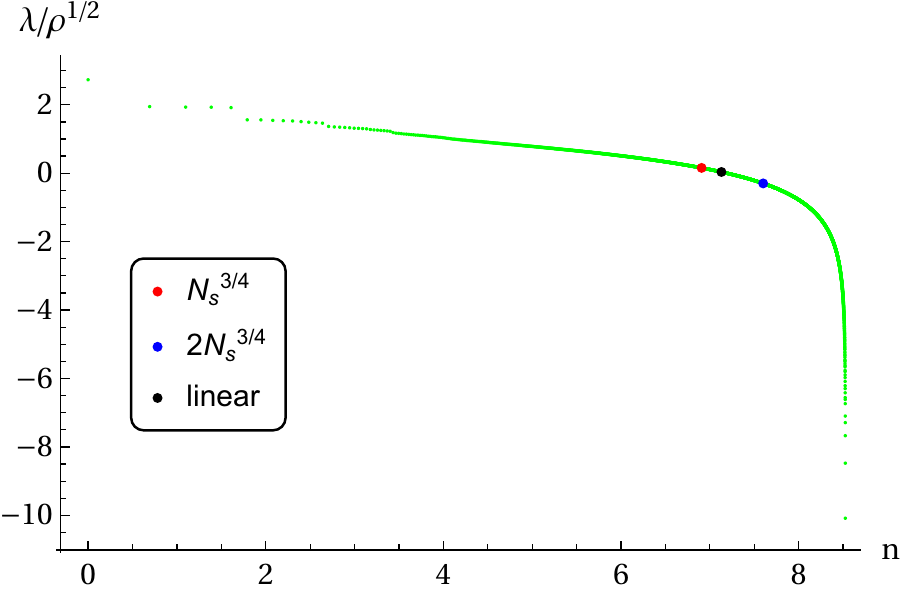}
  \caption{}
  \label{ds4spectrum}
  \end{subfigure}
  \begin{subfigure}[b]{0.5\textwidth}
  \includegraphics[width=\textwidth]{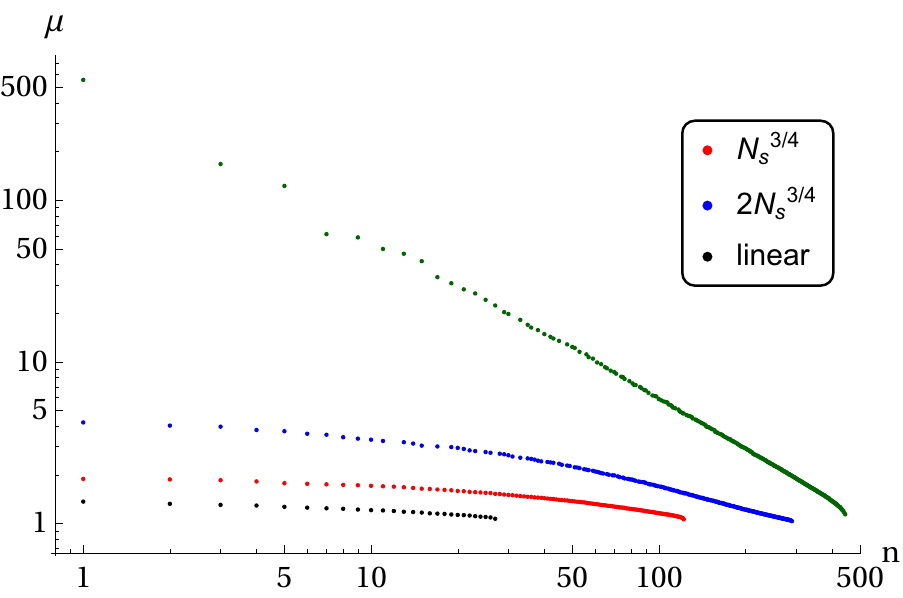}
  \caption{}
  \label{dS4 generalized spectrum}
  \end{subfigure}
  \caption{For $N=10k$ in $\deS_4$, (a) is the spectrum of $i\Delta$ with different truncations marked, and (b) is a plot of the
    solutions of the generalised equation \eqref{s4c} for these truncations.}
  \end{figure}

 In Figure \ref{ds4spectrum} we show the causal set SJ spectrum in the $\deS_4$ slab with the different truncations marked, and in Figure
 \ref{dS4 generalized spectrum} we show the generalised spectrum with and without these truncations. Again the broad features are the same -- the truncations lie in the linear regime of the SJ
 spectrum and drastically cut down both the magnitude and number of the generalised spectrum. However, the differences in the generalised spectrum post truncation are more
 marked in $d=4$.
 
 The area law for the SSEE for $\deS_4$ is given by
 \begin{equation}
   \ssee = a \sqrt{N} + b. 
   \end{equation} 
 With $h=1.2$, we expect $a \sim 0.17$ for the  Bekenstein-Hawking entropy for the $\deS$  horizon
 \eqref{4dcsentropy}.  In Figure \ref{ds4entropytrunc} we show the results for the SSEE. We note
 that interestingly, the scatter is far less than in $d=2$, which makes the results easier to interpret.

{ In all  cases we see that an area law and complementarity are compatible with the data, but that the 
 linear truncation scheme is also compatible with a volume law. 
 The  number truncations  $n_\mx=N^{\frac{3}{4}}$, and  $n_\mx=2 N^{\frac{3}{4}}$ give a much more convincing area law.   

A comparison with the Bekenstein-Hawking formula however shows that {\it all} the values of $a$ in Figure \ref{ds4num} exceed
 the expected value of $a=0.17$. So even though an area law is obtained, it is one that contains more entropy than
 expected. For $n_\mx=N^{\frac{3}{4}}$ the SSEE  is about $5$ times larger, and the difference is even greater for $n_\mx=2
 N^{\frac{3}{4}}$. Perhaps this is not surprising because we do not know the proportionality constant $\alpha$  in
 \eqref{eq:numtrunc}, although it is reasonable to expect an $\alpha$  of order $1$.  The linear truncation gives an SSEE that is closer to the Bekenstein-Hawking entropy, although it
again is in excess.}
 \begin{figure}[!h]
  \begin{subfigure}[b]{\textwidth}
  \includegraphics[width=0.5\textwidth]{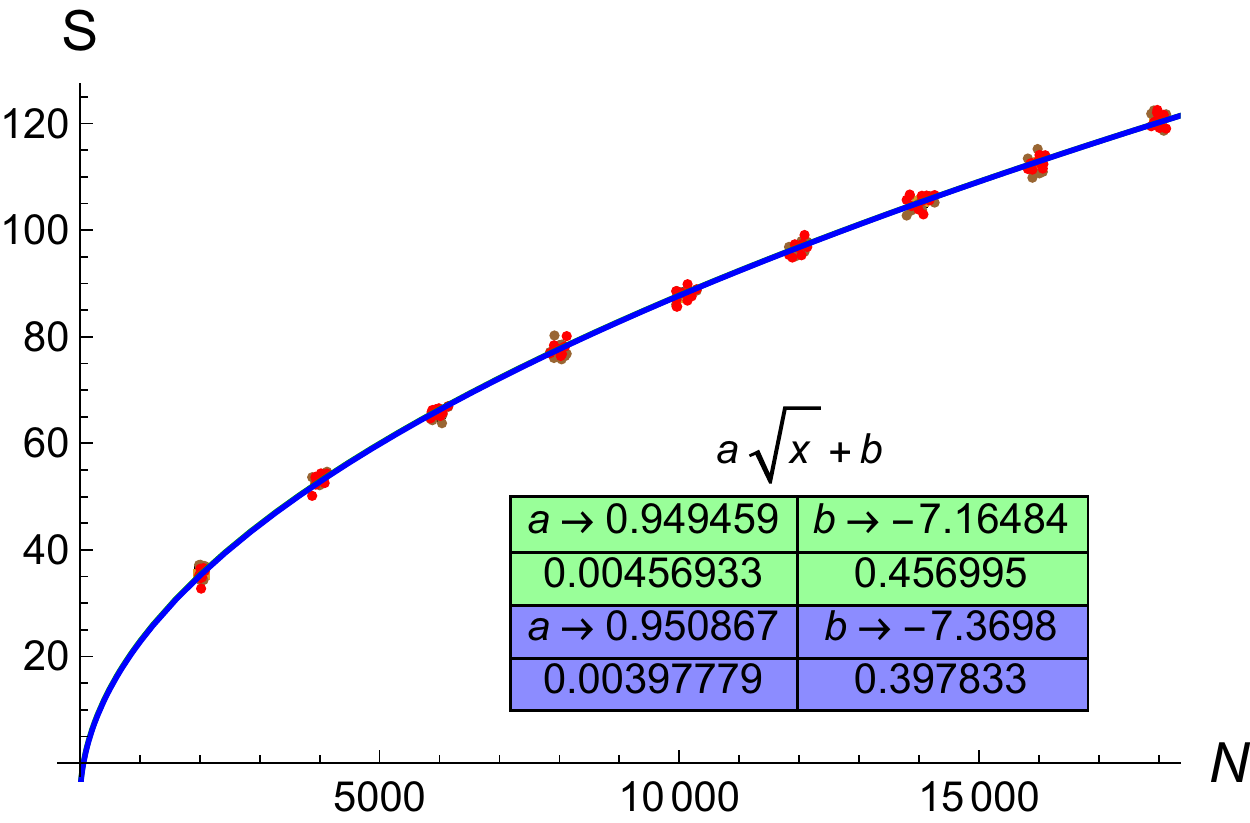}
  \includegraphics[width=0.5\textwidth]{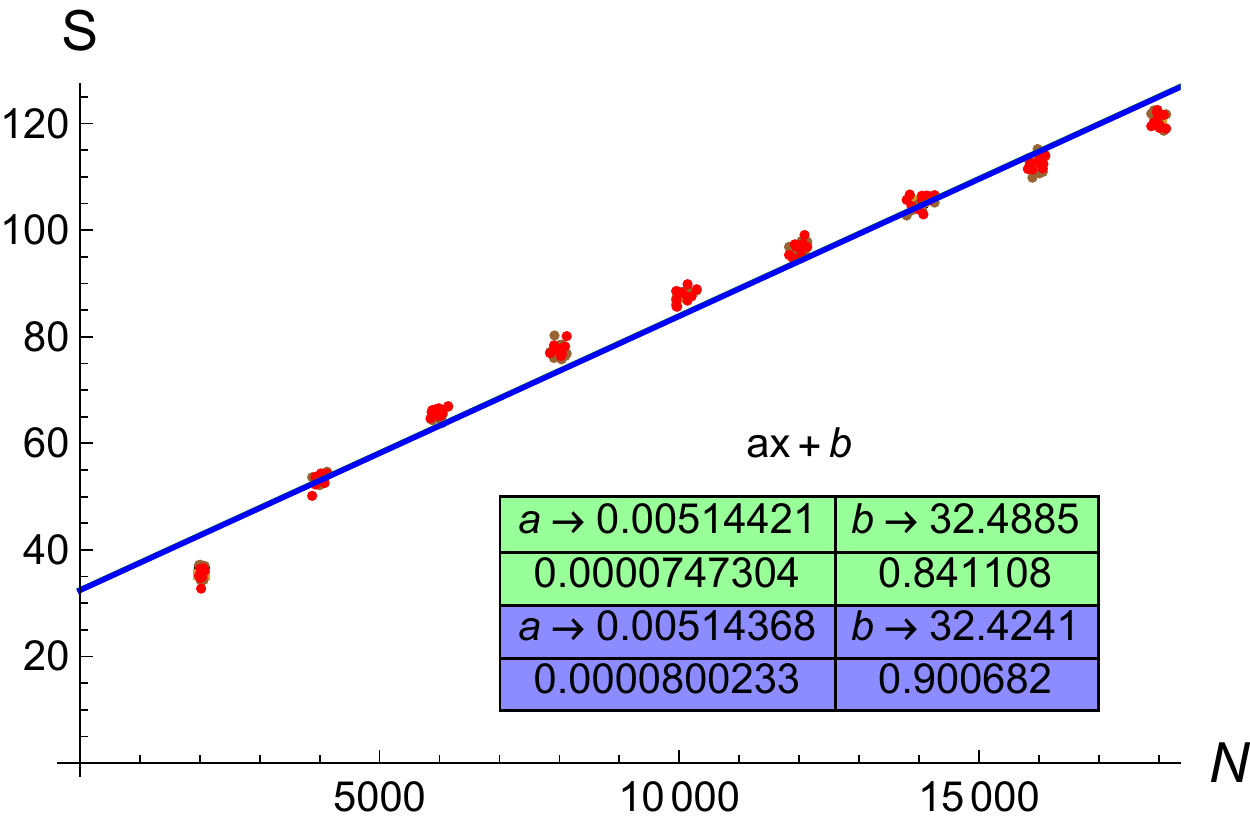}
  \caption{Number truncation with $n_\mx=N_s^{3/4}$}
  \label{ds4num}
  \end{subfigure}
  \begin{subfigure}[b]{\textwidth}
  \includegraphics[width=0.5\textwidth]{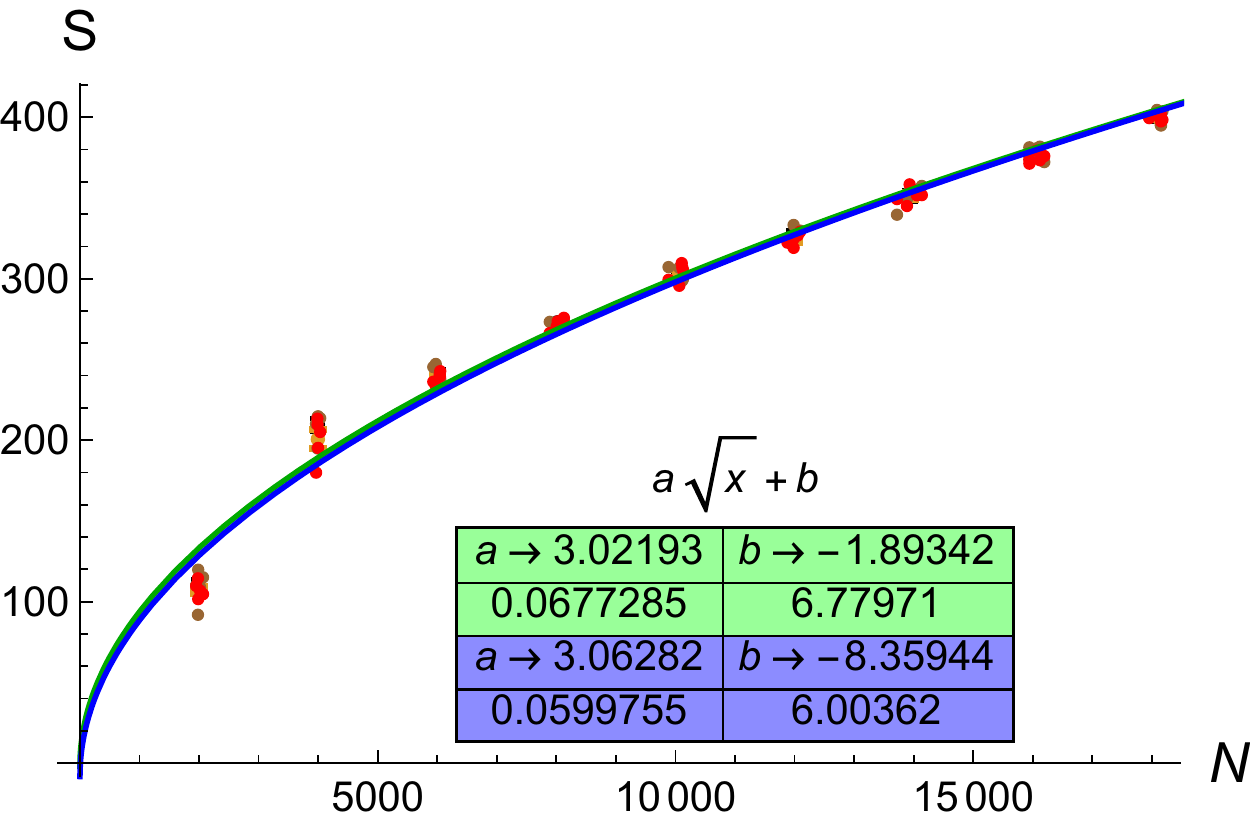}
  \includegraphics[width=0.5\textwidth]{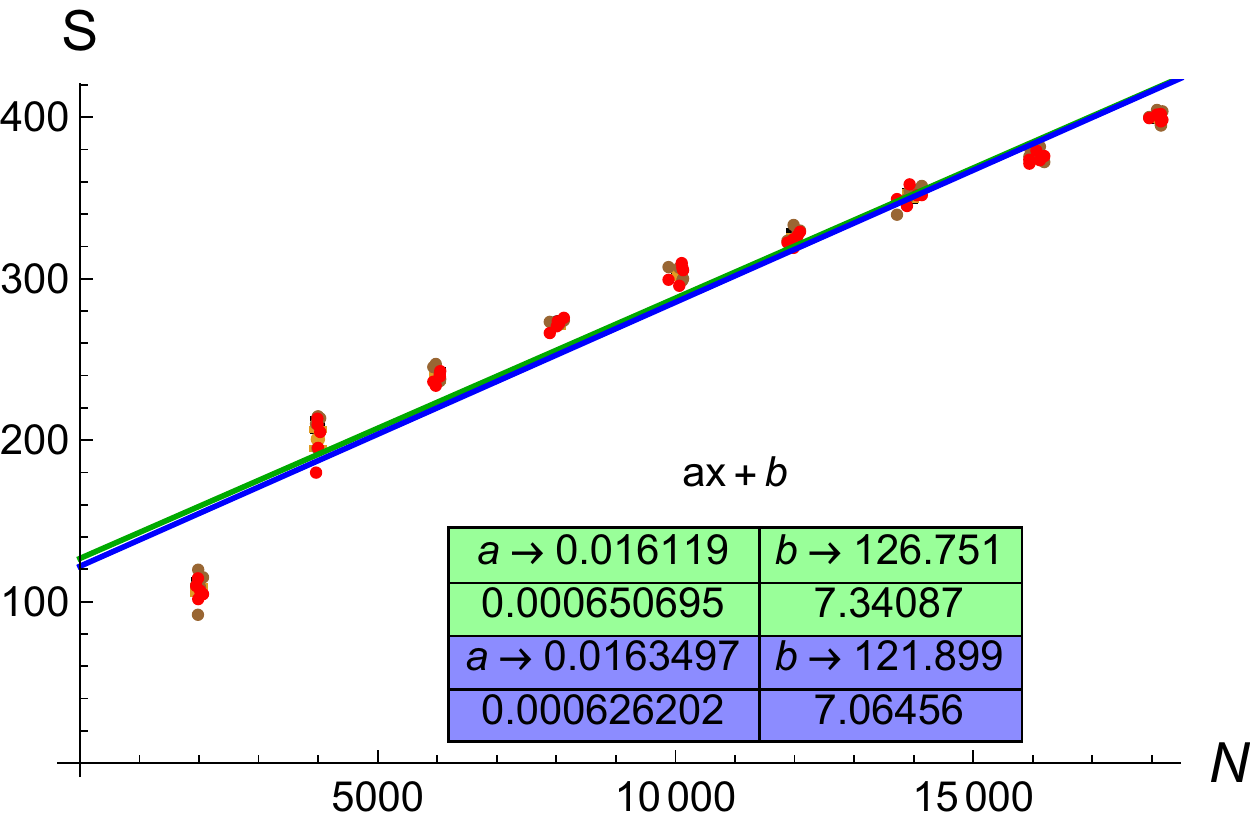}
  \caption{Number truncation with $n_\mx=2 N_s^{3/4}$}
  \label{ds4num2}
  \end{subfigure}
  \end{figure}
  \begin{figure}[!h]
\ContinuedFloat
  \begin{subfigure}[b]{\textwidth}
  \includegraphics[width=0.5\textwidth]{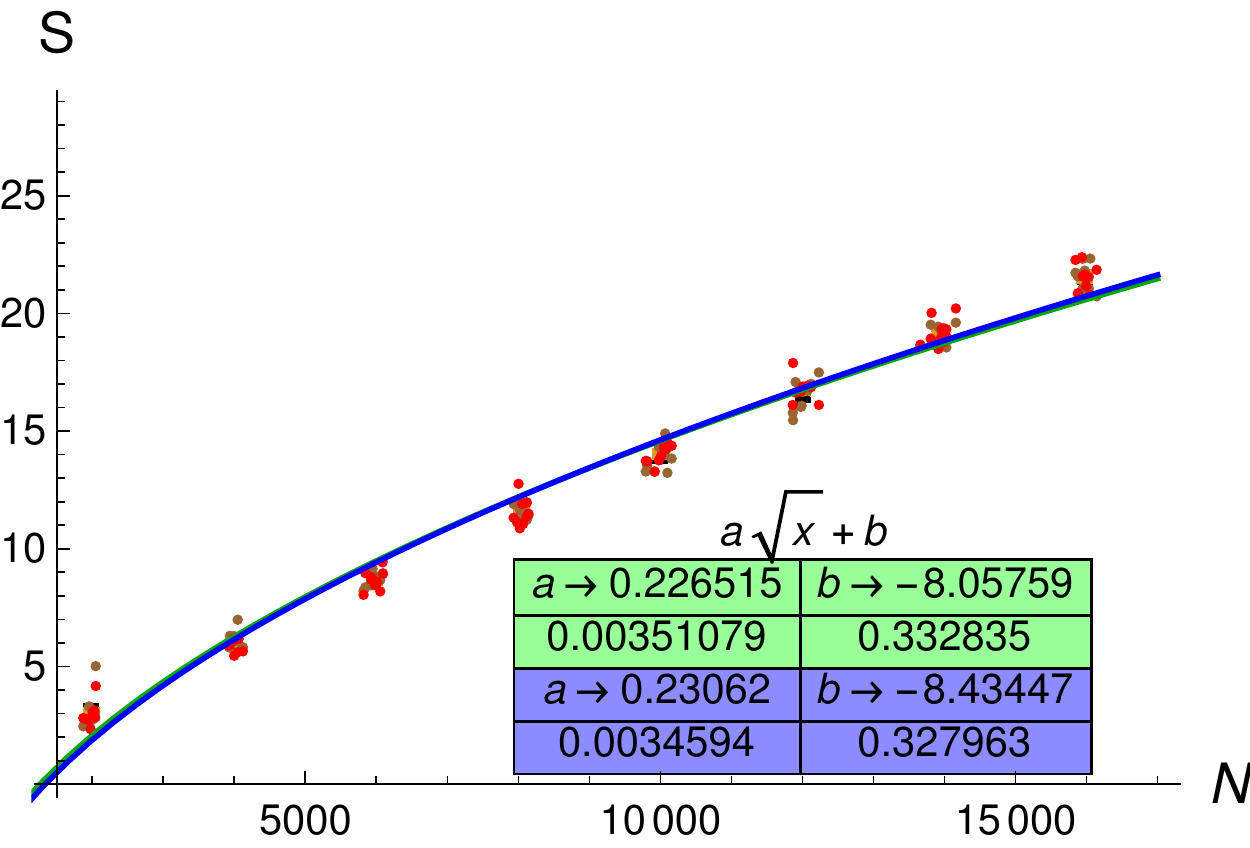}
  \includegraphics[width=0.5\textwidth]{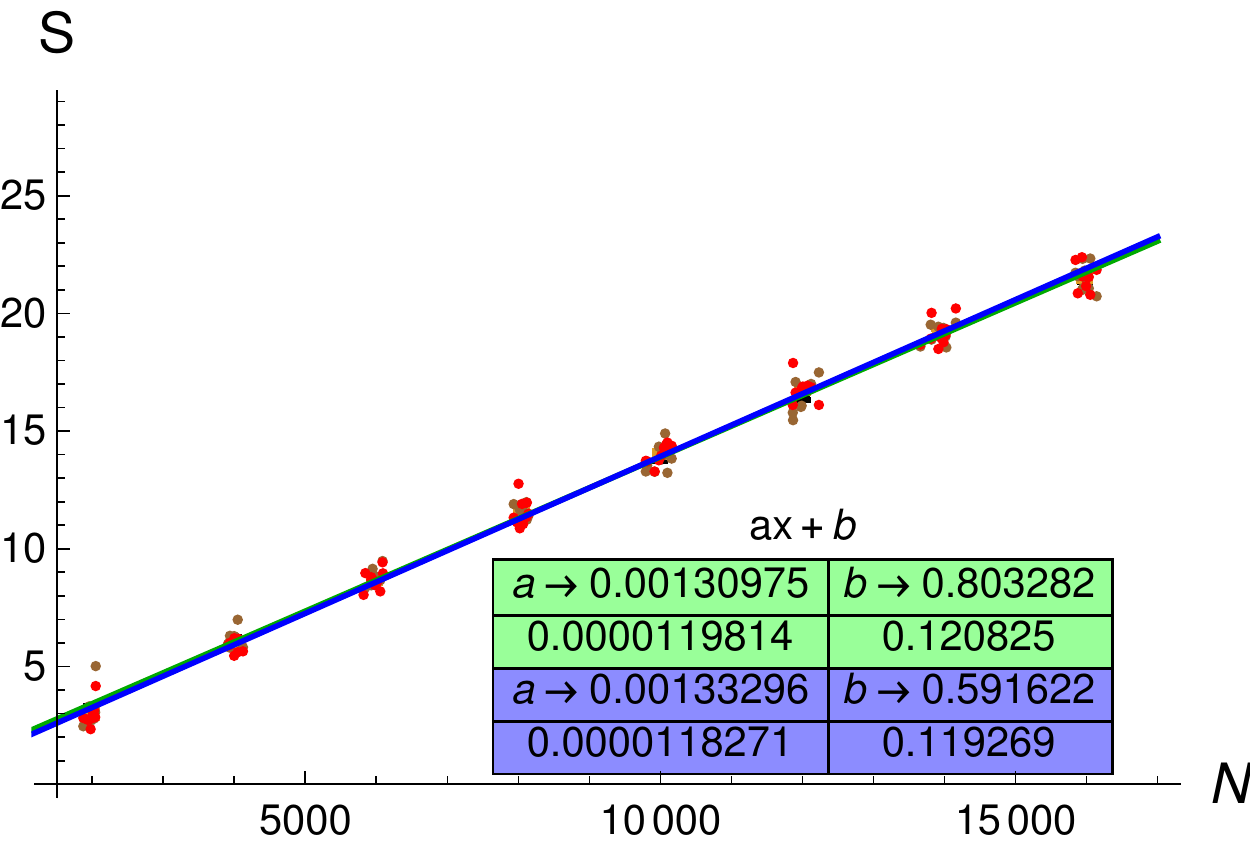}
  \caption{Linear truncation}
  \label{ds4linear}
  \end{subfigure}
  \caption{SSEE vs. $N$ with different choices of truncation in $\deS_4$.  The green and blue represent the data for
    the two Rindler-like wedges. A comparison of the two fits $a\sqrt{x}+b$ and $ax+b$ is shown on the left and the right
    for each choice of truncation.}
  \label{ds4entropytrunc}
  \end{figure}
\section{Discussion}
\label{discussion}

Numerical studies such as the one we have carried out in this paper have shown
that an excess SSEE due to the small eigenvalues of the SJ spectrum is a common occurrence, {and thus generically
  gives rise to a volume rather than an area law.} 
In this work we have presented evidence that the causal set SSEE for $\deS_{2,4}$  horizons satisfies a volume
rather than an area law, when the full causal set SJ spectrum is used.

On implementing {certain} double truncation schemes on the
SJ spectrum, inspired by the $\diam_L^2$ case,  we show that area laws can be  obtained, {which also, as expected, satisfy complementarity}.  In this sense, the properties of the causal set SSEE obtained in  \cite{Sorkin:2016pbz} for the nested
causal diamonds $\diam_\ell^2\subset \diam_L^2$ appear to be universal.

Out of the many truncation schemes explored in $\deS_2$, the number truncation $n_\mx=2 \sqrt{N}$ gave results most
  compatible with the Bekenstein-Hawking entropy, while for $\deS_4$, the linear truncation gave the best
  results. In both cases, the area law is satisfied up to the expected order of magnitude. There is a small
  over-estimation of the EE compared to the BHE in both cases, but we cannot attribute any physical significance to it
 since the exact value of the EE depends not only on the choice of truncation scheme but also on the parameters $\alpha$
  and $\delta$. For now it suffices that there are choices of these parameters which give us the correct order of magnitude estimate for
  the area law.   

In our investigations, several choices of these parameters have been scanned, with the values
 exhibited here being 
  the most optimal in terms of the area law and data compatible with complementarity. While the possibility always exists of further
 fine-tuning, or using  a different truncation scheme, this is not necessarily helpful without further physical
 understanding. One might of course also resort to the possibility that $N$ is not large enough and what we are seeing are 
finite size effects.  {We thus view our present} study {as} only the start of a more systematic {\sl analytic}
 understanding of the nature of the  causal set SSEE in curved spacetimes. 

There are   several distinctive features of the dS EE calculation which we now summarise.  To begin with, the SSEE
\eqref{s4c} does not specify a choice of vacuum. However, the SJ  vacuum prescription  is the only way we know how to define a vacuum in the
causal set (in the continuum it is a uniquely defined vacuum  in a finite spacetime
region)  and hence this is the choice we make. Importantly,  the causal set $\deS$   SJ vacuum
is distinct from the known Mottola-Allen vacua in the continuum as shown in 
\cite{Surya:2018byh}.  Thus, if we are to take the causal set as more fundamental than the continuum, this suggests that  the ``correct''
QFT vacuum for calculating the SSEE  is not one of the Mottola-Allen vacua, but rather one with modified  UV properties. 

The truncation procedure, while essential for an area law, remains poorly understood. The fact that the SSEE obeys a volume law without
truncation seems to arise from the non-locality of the causal set. As discussed in \cite{Eisert:2008ur} systems with
long-range order exhibit volume rather than area laws. The non-locality in a causal set, which enables an element near
the past boundary to be linked to one near the future boundary, is fundamental to the discrete-continuum
correspondence. Localising influences near sets of measure zero, even if they are genuine horizons, is not commensurate
with this feature. Thus, a volume law seems particularly convincing in causal set theory. However, the lack of complementarity due to a volume law could also mean that the entanglement entropy is not bipartite, or that the entropy we are computing is dominated not by the entanglement entropy but by some other form of entropy (like the entropy of coarse graining). These issues need to be investigated further. Since area laws are a
fundamental feature of General Relativity, which causal set theory must approximate, locality must be emergent, and with it, an area
law for the SSEE.

Truncation also throws up new problems, namely the possibility of causality violation, as for  example in the nested
causal diamonds in $\mink^2$.  Here, the condition $i\Delta(x,x')=0$ for  spacelike separated points $x$ and $x'$ does not always hold for
the truncated $i\Delta$. It can be shown that this acausality averages to zero over multiple sprinklings, but this
nevertheless  points to the need for a deeper understanding of the truncation process. 

A possible way in which truncation may be avoided is to note that there could be significant deviations from
manifold-likeness in the causal set at smaller scales. The  correct UV completion of spacetime is likely not
manifold-like at all and could significantly  change the SJ spectrum and in turn the SSEE. This  conjectured  UV
completion may be the missing ingredient in our discussion, but we are far from an understanding of what this might be.

As we have seen and as shown in \cite{Sorkin:2016pbz},  while the causal set offers a ready covariant spacetime
cutoff, the recovery of an area law for the SSEE in $\deS$, though possible, is not straightforward. Suggestions in \cite{Belenchia:2017cex} for  a deeper understanding from an Algebraic  Quantum
Field Theory perspective need to explored further. Recently the continuum $\deS$  SJ spectrum in the slab has been
found analytically \cite{AMSS}, and offers us a possible route to calculating the  continuum $\deS$  SSEE, and hence
finding a more physically motivated truncation. Future work in this direction should hopefully shed some light on the
questions that we have raised in this work.  
\newline

\bf Acknowledgements: \rm We thank Fay Dowker for helpful discussions. YY acknowledges  financial support from Imperial College
London through an Imperial College Research Fellowship grant, as well as support from the
Avadh Bhatia Fellowship at the University of Alberta. SS  is supported in part by a Visiting Fellowship at the Perimeter Institute. Research at Perimeter Institute is supported in part by the Government of Canada through the Department of Innovation, Science and Economic Development Canada and by the Province of Ontario through the Ministry of Colleges and Universities.

\begin{appendices}
 
\section{Causal Sets and Sprinkling}
\label{cs.app}
 
 Causal sets are proposed discrete underpinnings of spacetime. They are partially ordered sets consisting of a set of spacetime elements and the ordering relation among them. 
 The causal set elements $x\in\mathcal C$ and their ordering relation (which is the causal precedence relation $\preceq$) 
satisfy a number of conditions: 
$\forall x\in \mathcal C$, $x \preceq x$ (reflexivity);
$\forall x,y\in \mathcal C$, $x\preceq y\preceq x$ implies $x=y$ (antisymmetry);
 $\forall x,y,z\in \mathcal C$, $x\preceq y\preceq z \implies x\preceq z$ (transitivity);
$\forall x,y\in \mathcal C$,
$|I(x,y)|<\infty$, where $|\cdot|$ denotes cardinality and $I(x,y)$
is the causal interval defined by $I(x,y):=\{z\in \mathcal C|x\preceq z\preceq y\}$, (local finiteness).
 
\begin{figure}[!h]
 \centering
	\includegraphics[width=0.8\textwidth]{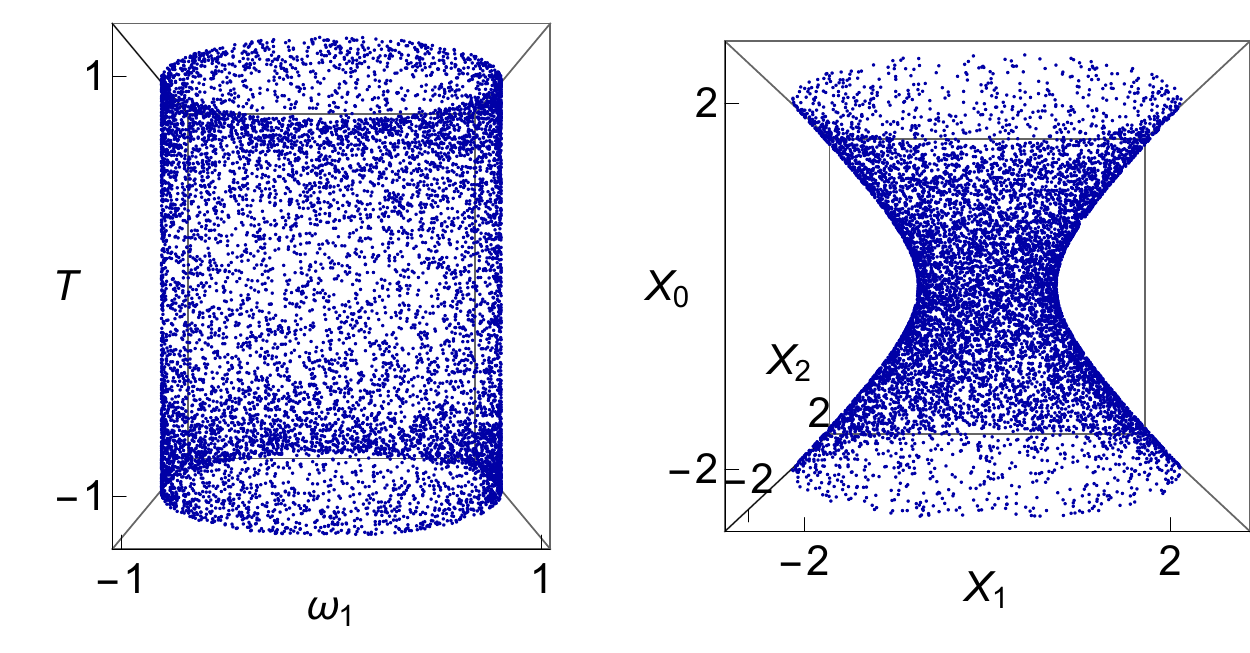}
 	\caption{A sprinkling of $N=10000$ elements into the metric of \eqref{confmetric} for the time interval $-1.2<T<1.2$.}
 	\label{sprinkling}
 \end{figure}

 Ultimately the causal set dynamics will dictate which causal sets are produced and how. At present this dynamics is a work in progress. Meanwhile a causal set can be produced through the process of sprinkling: randomly placing points  in a Lorentzian manifold following a  Poisson distribution. Sprinklings in $\deS$   are shown in Figure \ref{sprinkling}.

 A common representation of a causal set is by a type of adjacency matrix called the  causal matrix $C$. The causal matrix is defined as
\begin{equation}
C_{xy}=\begin{cases}
1 & \text{for $x\preceq ~y$}\\
0 & \text{otherwise}
\end{cases}
\label{cmatrix}
\end{equation}

We have used the notation that the indices $xy$ correspond to the matrix entry relating causal set elements $x$ and $y$.
 
 Another useful representation of a causal set is using the  link matrix $L$. The link matrix is defined as
\begin{equation}
L_{xy}=\begin{cases}
1 & \text{for $x\preceq\!\!* ~y$}\\
0 & \text{otherwise}
\end{cases}
\label{lmatrix}
\end{equation}
where we have introduced a link $\preceq\!\!*$ as a nearest neighbor relation such that $x \preceq y$ but with no $z \in C$ as an intermediate relation $x \preceq z \preceq y$. The link matrix $L$ is also sometimes labelled as $L_0$. The retarded Green functions discussed in the next appendix are expressed in terms of the causal and link matrices.

\section{The SJ Vacuum and Green Functions}
\label{green.app}
 
Entanglement entropy is defined relative to a pure state. When a quantum field is in a pure state, its entropy of entanglement is zero. If the geometry in which the field resides is split into two complementary subregions, the quantum field's restriction to either subregion is no longer pure (it becomes mixed) and produces a non-zero EE. The pure state of SSEE is given by the Sorkin-Johnston (SJ) vacuum state, namely the SSEE vanishes when the quantum field is in the SJ state. We review the definition of the SJ vacuum state for a free scalar quantum field in this appendix. A key ingredient in the construction of this state is the retarded Green function. We also review the definition of this Green function in causal set theory. 

The SJ prescription uses the Pauli-Jordan or spacetime commutator function 
\begin{eqnarray}
i \Delta  (x,x') &=&\langle 0| [ \phi  (x),\phi  (x'    )   ]|0  \rangle\nonumber\\
 &=& {W } (x,x ')-{W }(x',x),
 \label{idelta}
\end{eqnarray}
where $x$ and $x'$ are points in {\sl{spacetime}}, and $W$ is the Wightman or two-point correlation function. 
$i \Delta $ is a c-number and \eqref{idelta} is therefore  independent of the state in which it is computed. This is in contrast to $W$ which is state dependent and in fact can be used to define the state. An alternative representation of $\Delta$  is the following one in terms of the  retarded and advanced Green functions 
\begin{equation}
    \Delta(x, x')=G_{R}(x, x')-G_{A}(x, x'),
\label{delta}
\end{equation}
where $G_{R}$ and $G_{A}$ both satisfy $(\Box-m^{2})G_{R,A}=-\delta^{d}(x-x')/\sqrt[]{-g}$ and $G_A=G_R^T$. 

 In $d=2$ and for any causal set approximated by flat or curved continuum geometries, the retarded Green function is
 
 \begin{equation}
 G_{R}^{(2)}=\frac{1}{2} C (\mathbb{I}-\frac{m^2}{2\rho} C)^{-1},
 \end{equation}
 
 where $C$ is the causal matrix and $\mathbb{I}$ is the identity matrix.
 
 In $d=4$, and for causal sets approximated by Minkowski spacetime or the $\deS$  case with coupling $\xi$, the retarded Green function is \cite{NomaanX2017}
 \begin{equation}
 G^{(4)}_{R}(x,x')\equiv\sum^{N}_{k=0} a^{(k+1)} b^k L_k(x,x')=a L_0 (\mathbb{I}-b a L_0)^{-1},
 \label{dsgret}
 \end{equation}
 
 where $a=\frac{1}{2\pi}\sqrt{\frac{\rho}{6}}$,  $b=-\frac{1}{\rho}(m^2+(\xi-\frac{R}{6}))$, $L_k$ is the product of the link matrix \eqref{lmatrix} with itself $k$ many times, $R$ is the scalar curvature, $m$ the mass\footnote{For a detailed discussion of the concept of mass in $\deS$   see \cite{Garidi2003}.}, and $N$ is the size of the causal set.

Since $i\Delta$ (or $G_R$, from which one obtains $i\Delta$ through \eqref{delta})  is the starting point of the SJ  prescription, this makes the prescription manifestly  covariant. In a region of spacetime $\mathcal{R}$, $i \Delta $ defines an integral operator over the Hilbert space of square integrable functions ${\mathcal{L}}^{2}(\mathcal{R})$. Under  suitable conditions (which are satisfied in finite-volume regions of spacetime such as the ones we consider in this paper) and in  globally hyperbolic spacetimes (or subregions), $i\Delta$ is self-adjoint and can be decomposed into its positive and negative eigenspace.
 
Let $u_{k},v_{k}$ be the normalised positive and negative eigenfunctions of $i\Delta$ respectively, and $\pm\lambda_k$ their eigenvalues, such that
\begin{equation}
  \int_{ }^{ } dV' i\Delta(x,x') u_{k}(x')=\lambda_{k} u_{k}(x),
\end{equation}
\begin{equation}
 \int_{ }^{ } dV' i\Delta(x,x') v_{k}(x')=-\lambda_{k} v_{k}(x),
\end{equation}
where $dV$ is the invariant volume element of the spacetime and $\lambda_{k}>0$, then $i\Delta$ can be expressed in its eigenbasis as the expansion
\begin{equation}
i \Delta  (x,x '  )= \sum_{   k}\{\lambda_{k} u_{k}(x)u_{k}^{\dagger}(x')-\lambda_{k} v_{k}(x)v_{k}^{\dagger}(x')\}.
\label{expansion}
\end{equation}
Finally, the SJ prescriptions is to define the SJ Wightman function ${W}_{SJ}$ as a restriction of the expansion \eqref{expansion} to its positive part
\begin{equation}
{W}_{SJ}(x,x')\equiv  \text{Pos}(i\Delta)=\sum_{  k}^{ }\lambda _{k}u_{k}(x)u_{k}^{\dagger}(x').
\end{equation}
Every Wightman function  defines a vacuum state, so the SJ Wightman function  defines the SJ vacuum state of the theory. The field operator $\phi(x)$ in terms of these eigenfunctions is $\phi(x)=\sum\limits_{k}\sqrt{\lambda_{k}}\{a_{k}u_{k}(x)+a^{\dagger}_{k}u^{*}_{k}(x)\}$, but we will not require the field and will only need to work directly with $W$ in \eqref{s4} or \eqref{s4c}. Since we consider Gaussian theories, $W$ contains all the information in the theory. For some studies of the SJ vacuum see  \cite{Afshordi:2012ez,Surya:2018byh, abis,Buck:2016ehk}.

From its definition it is evident why the SJ state is a pure state for the SSEE. Since $W_{SJ}$ is the positive eigendecomposition of $i\Delta$, the eigenvalues of $W_{SJ}$ coincide with the positive eigenvalues of $i\Delta$. Therefore in \eqref{s4}, the solutions $\lambda$ will be solely $1$'s and $0$'s. Upon inserting these in $\lambda\ln\lambda$ and summing them in \eqref{s4} we get zero contribution to the entropy.\footnote{This is the case because both $0\ln 0=0$ and $1\ln 1=0$.}

The SJ vacua in many cases agree with conventional choices of vacua (for example in static spacetimes
\cite{Afshordi:2012jf}) but are more general.  In \cite{Surya:2018byh} the SJ vacua in $\deS$   were studied in
detail. Among the cases that were studied were those that we consider in this paper.  These include causal set
discretisations of  $\deS_{2,4}$  with a range of masses $m \geq 0$ for  the free scalar field. In some cases, such as
the massive theory in $\deS_2$, the SJ vacuum coincided with other known vacua known as the Motolla-Allen $\alpha$-vacua
\cite{Allen:1985ux, mottola}. In general, however, it was found that the vacua did not coincide with the continuum
$\alpha$-vacua. The causal set SJ vacuum in $\deS_4$ for example, while de Sitter   invariant, does  not resemble any known vacua from the continuum. Also, interestingly, the causal set SJ vacuum is well-defined for all masses and couplings, including the minimally coupled massless case which is typically considered ill-defined \cite{ Allen:1985ux,Allen:1987tz}.
In this paper we use the SJ vacua found in \cite{Surya:2018byh} as our starting point, i.e., as the pure states of our SSEE. We use $W_{SJ}$ in \eqref{s4c} and we use the causal set retarded Green function \eqref{dsgret} to define $i\Delta$ in \eqref{s4c}.

\section{Nested Causal Diamonds in $\mink^4$}
\label{mink.app}

\begin{figure}[!h]
	\centering
	\includegraphics[width=0.7\textwidth]{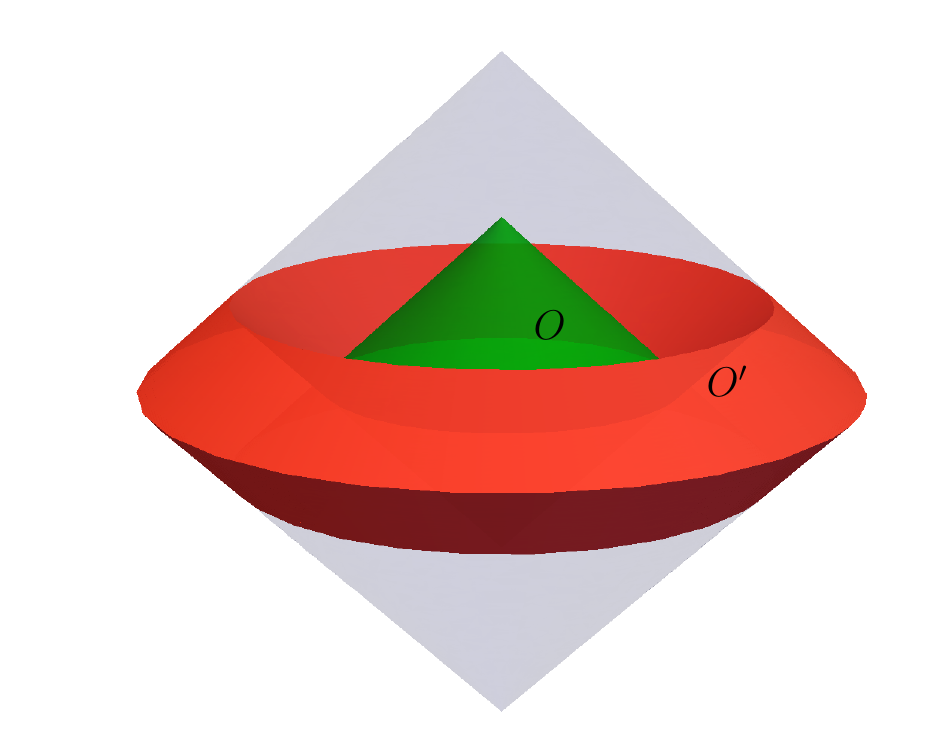}
	\caption{A nested causal diamond $O$ and its complement $O'$ in $3d$.}
	\label{fig:4ddiamond}
\end{figure} 

Here we present some results for nested causal diamonds in $\mink^4$ which is a non-trivial extension of the $\mink^2$
case. 

We consider a similar set up to $\mink^2$, namely nested causal  diamonds $\diam^4_\ell \subset \diam_L^4 \subset \mink^4$. The causal complement $O' \subset \diam_L^4$ of $\diam_\ell^4$ is connected and is the domain of dependence of a $d=3$ open ball with a concentric spherical hole. We show this connectivity in Figure \ref{fig:4ddiamond}, where one of the dimensions has been suppressed.

In $\diam^4$, the entropy-area relation $S=A/4$ of \eqref{sh} is  
\begin{equation}
S= \pi r^2 , 
\end{equation}where $r$ is the radius of the smaller diamond.
The expression in the causal set, \eqref{eq: causalsetentropy}, is
\begin{equation}
S^{(c)}= \sqrt{\frac{3 \pi}{2}} (r/R)^2 \sqrt{N} ,
\label{4dcsentropymink} 
\end{equation}
where $R$ is the radius of the larger diamond. In our simulations we set $r/R=0.6$, therefore we expect $S \approx 0.78\sqrt{N}$. 

We consider a number truncation with $n_\mx = N^{3/4}$ in all regions, as well as a number truncation with $n_\mx = N^{3/4}$ in $\diam^4_L$ and $\diam^4_\ell$ while $n'_\mx = 2 N^{3/4}$ in the complement of $\diam^4_\ell$. The motivation for the factor of $2$ in the latter number truncation is that the relative spatial volume of the subset of the $t=0$, time-symmetric Cauchy slice that lies in the complementary region is around twice as large as the subset that lies in $\diam^4_\ell$. We also consider the linear truncation with $m'=-0.25-|\epsilon|$ and $\epsilon=0.05$ (or $\delta=0.2$) in all regions.
In the simulations, we consider $\langle N \rangle$ values ranging from $4000$ to $18000$. For each $\langle N \rangle$ we consider 5 realisations. 

\begin{figure}[!h]
	\begin{subfigure}[b]{0.55\textwidth}
		\includegraphics[width=\textwidth]{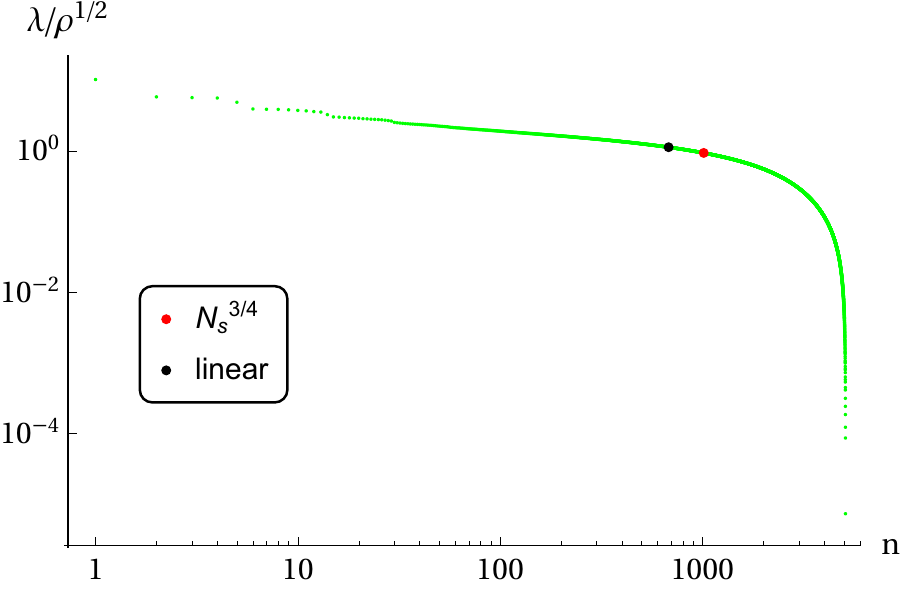}
		\caption{}
		\label{4dmink spectrum marked}
	\end{subfigure}
	\begin{subfigure}[b]{0.55\textwidth}
		\includegraphics[width=\textwidth]{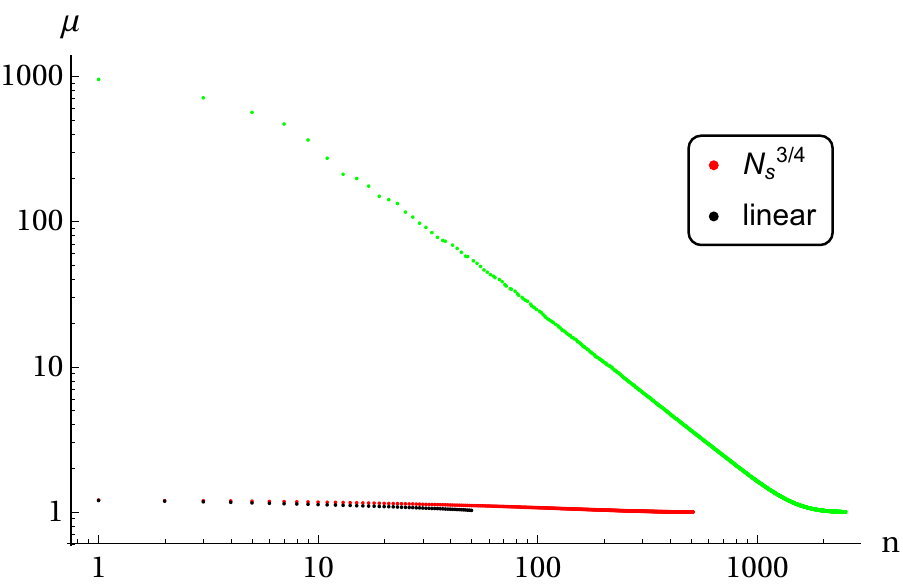}
		\caption{}
		\label{4dmink generalized spectrum}
	\end{subfigure}
	\caption{For $N=10k$, (a) is the spectrum of $i\Delta$ with different truncations marked, and (b) is a plot of the solutions of the generalised equation \eqref{s4c} for these truncation schemes.}
\end{figure} 
\begin{figure}[!h]
	\centering
	\includegraphics[width=0.6\textwidth]{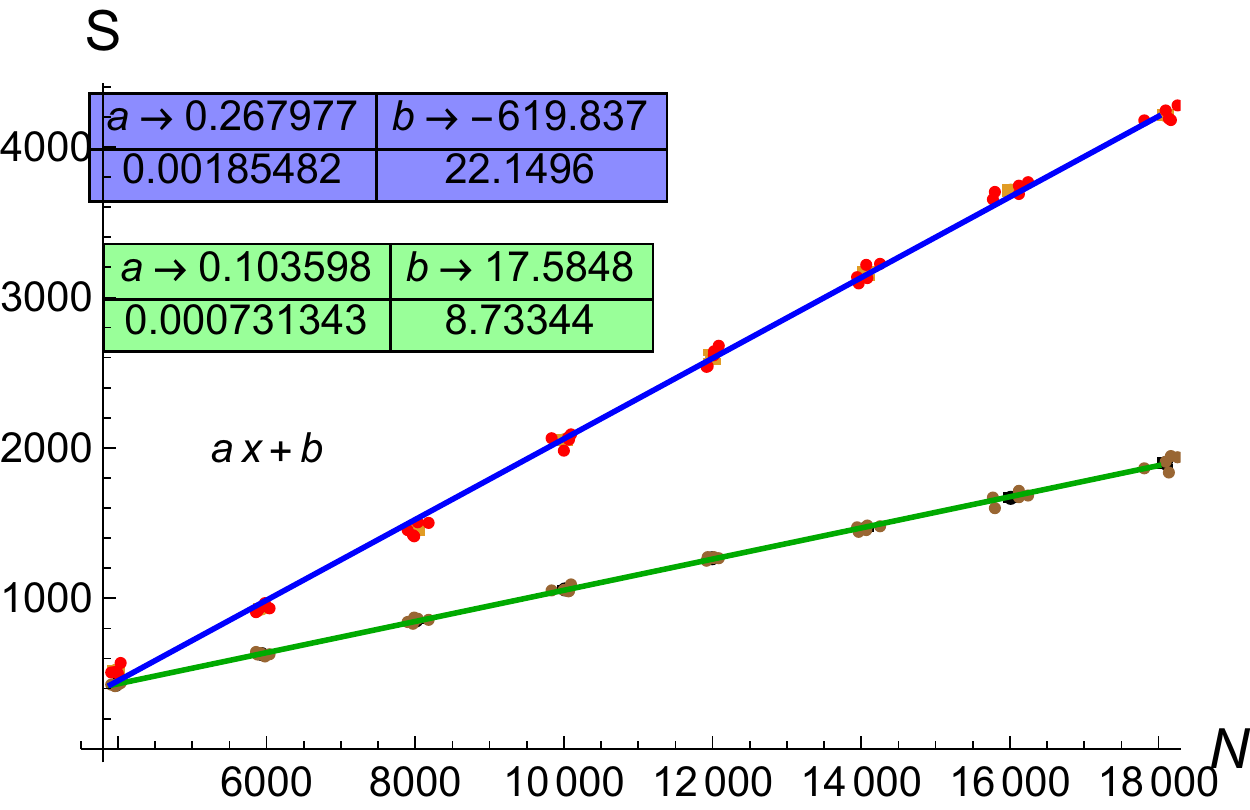}
	\caption{SSEE vs. $N$ without truncation. Green represents the data for $\diam^4_\ell$ and blue represents its complementary region. The best fits are shown. }
	\label{4dmink entropy notrunc}
\end{figure}
In Figure \ref{4dmink spectrum marked} we show the SJ spectrum for $\diam^4_L$, and where the truncations we consider lie. We also show the SSEE eigenvalues $\mu$ in Figure \ref{4dmink generalized spectrum} for one realisation.
The causal set SSEE without truncation is shown in Figure \ref{4dmink entropy notrunc} and can be seen to obey a spacetime volume law as anticipated. 
\begin{figure}[!h]
	\begin{subfigure}[b]{\textwidth}
		\includegraphics[width=0.5\textwidth]{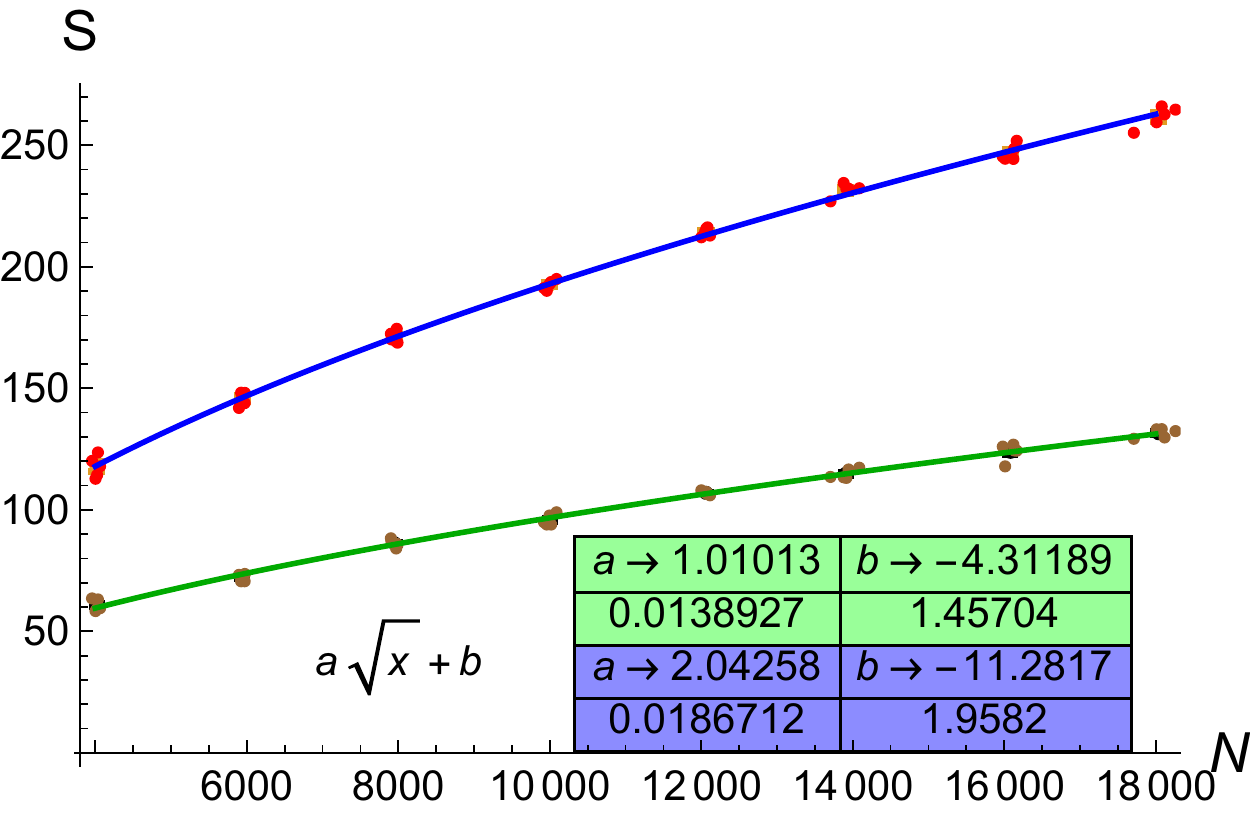}
		\includegraphics[width=0.5\textwidth]{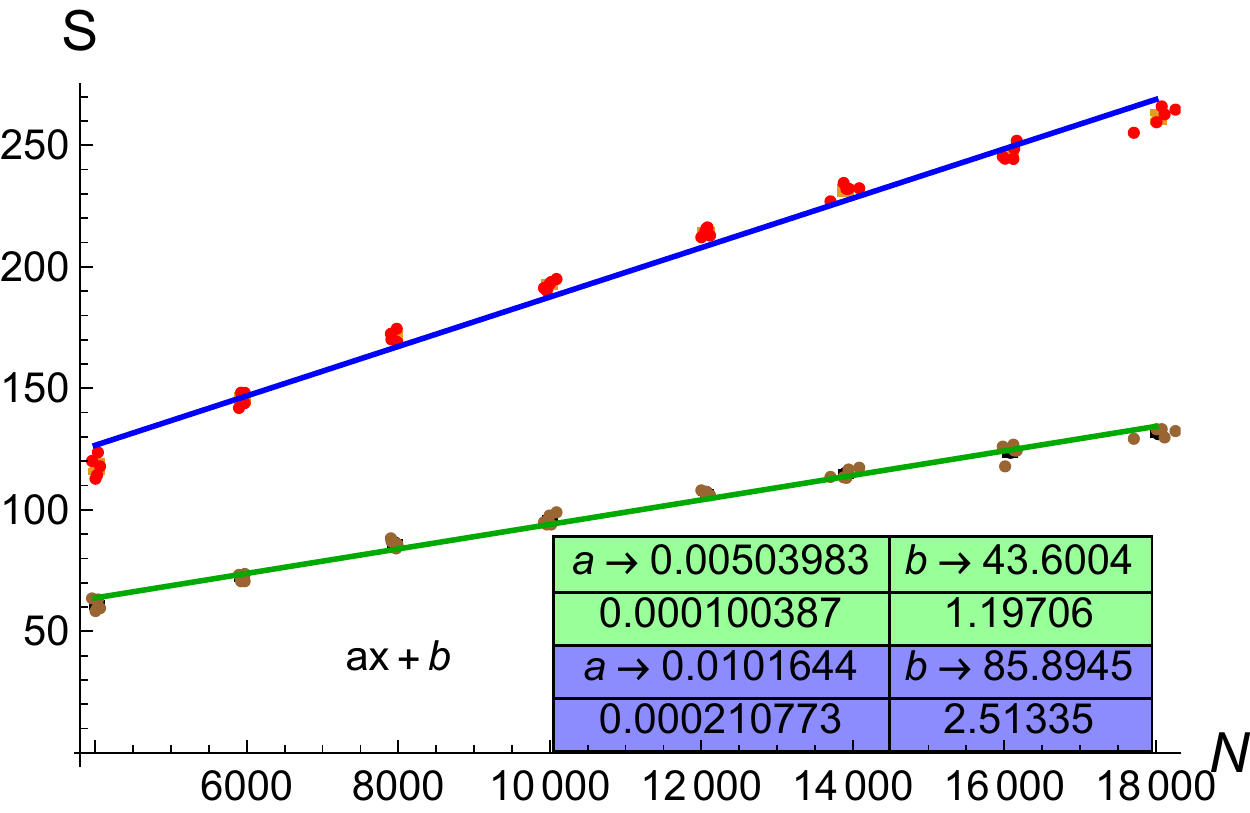}
		\caption{Number truncation with $n_\mx=N_s^{3/4}$}
		\label{4dmink entropy num nofactor}
	\end{subfigure}
	\begin{subfigure}[b]{\textwidth}
		\includegraphics[width=0.5\textwidth]{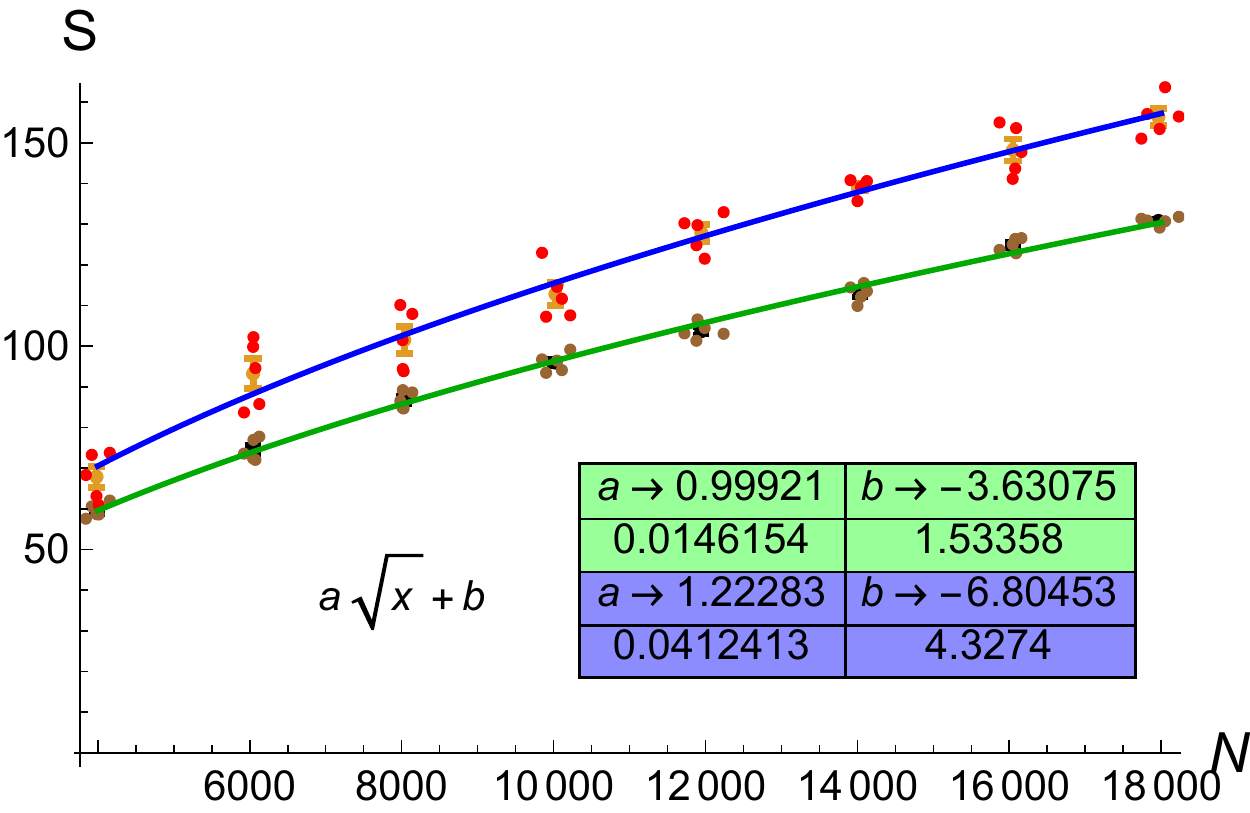}
		\includegraphics[width=0.5\textwidth]{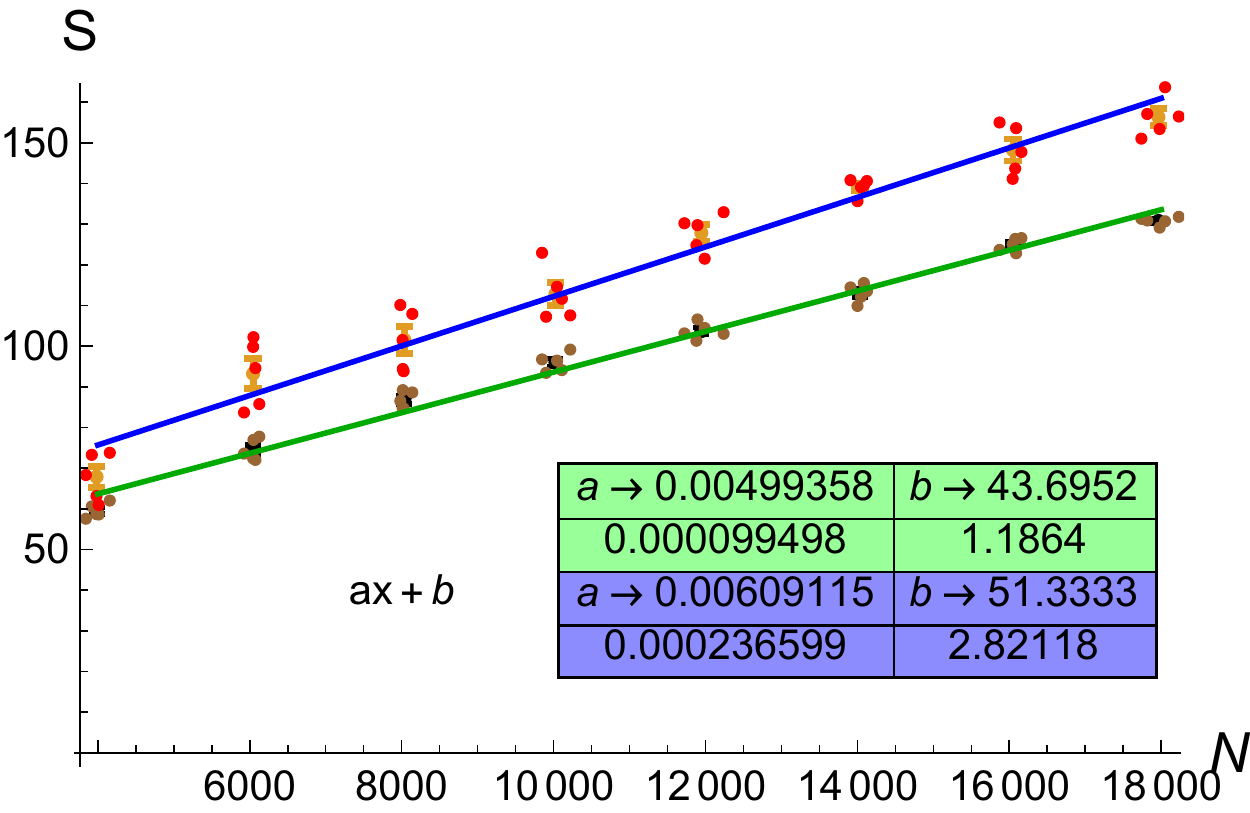}
		\caption{Number truncation with $n'_\mx=2 \,n_\mx=2N_s^{3/4}$}
		\label{4dmink entropy num factor2}
	\end{subfigure}
\end{figure}
\begin{figure}[!h]
	\ContinuedFloat
	\begin{subfigure}[b]{\textwidth}
		\includegraphics[width=0.5\textwidth]{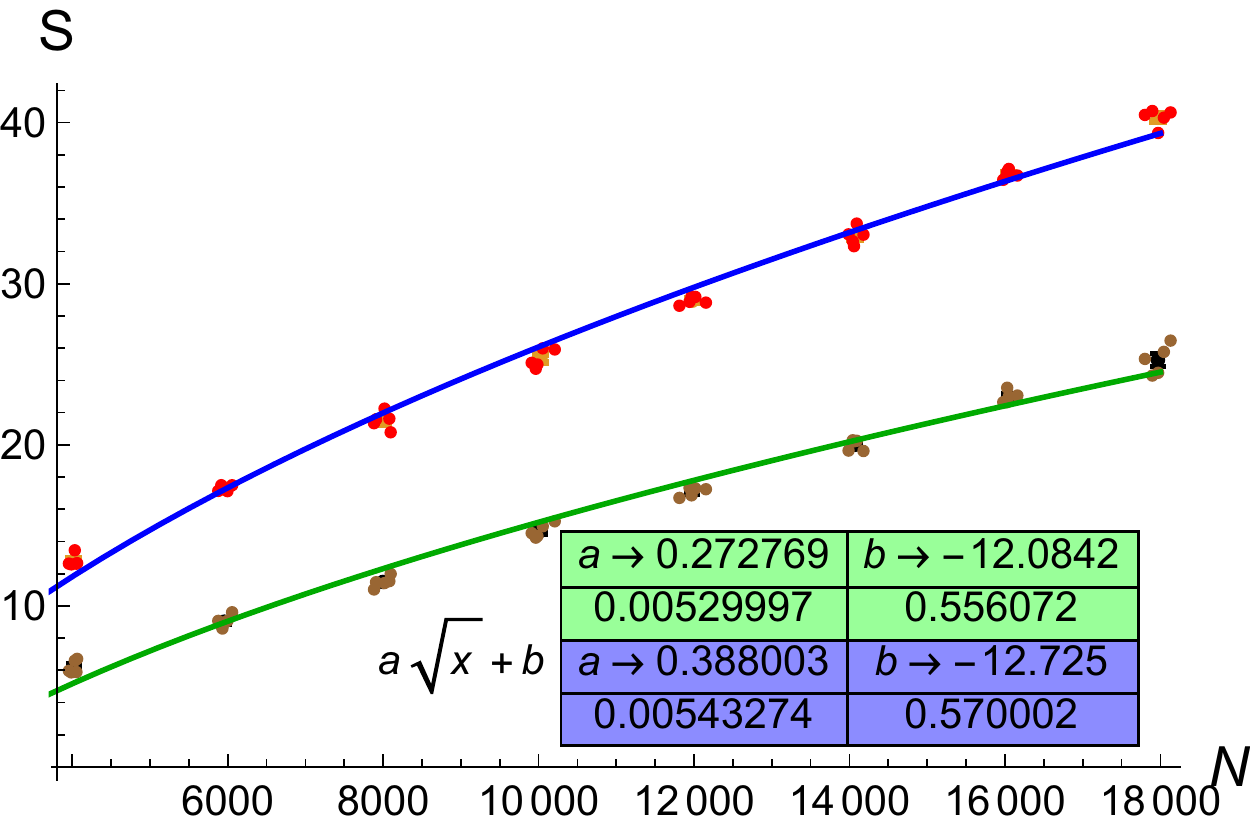}
		\includegraphics[width=0.5\textwidth]{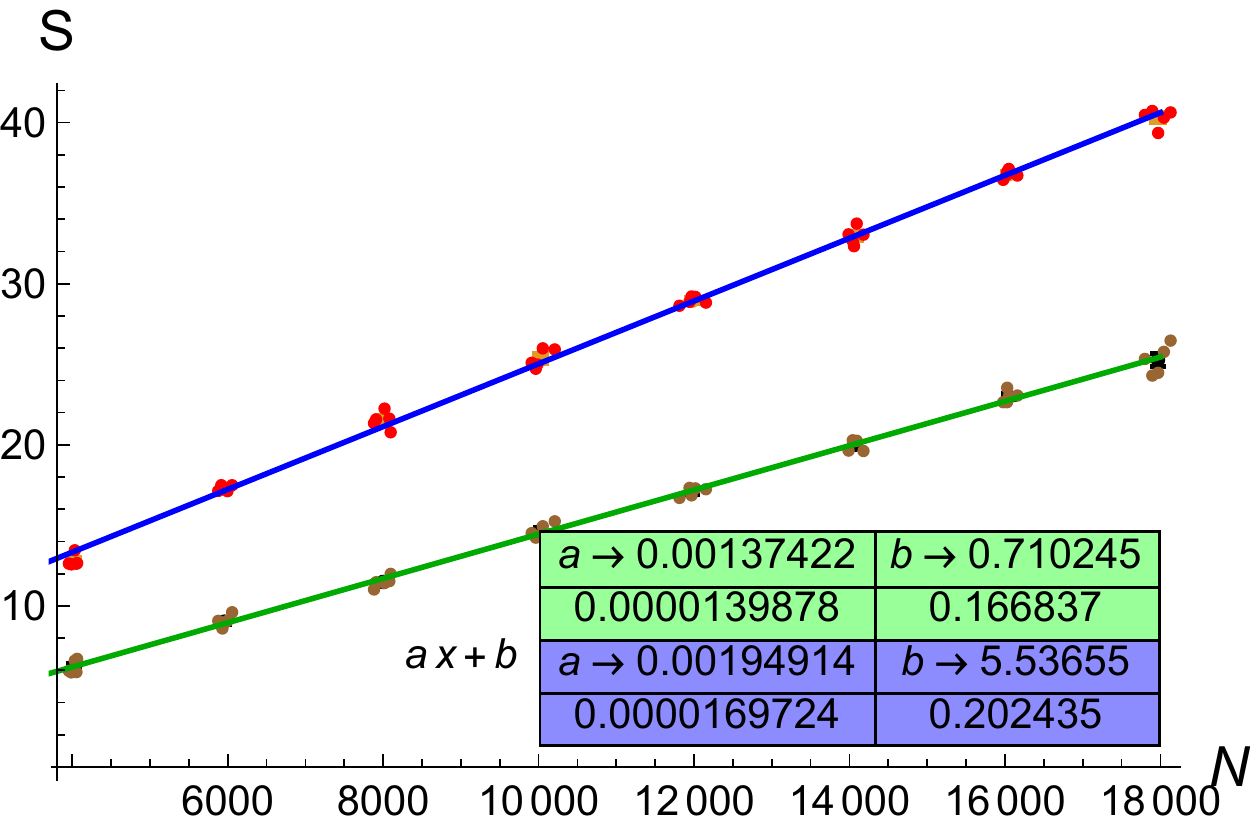}
		\caption{Linear truncation}
		\label{4dmink entropy dynamic linear}
	\end{subfigure}
	\caption{SSEE vs. $N$ with different truncations. Green represents the data for $\diam^4_\ell$ and blue represents its complementary region. A comparison of the two fits $a\sqrt{x}+b$ and $ax+b$ is also shown. Here $n'_\mx$ is the number truncation in the region complementary to $\diam^4_\ell$.} 
	\label{4dmink entropy trunc}
\end{figure}

Next, we show in Figure \ref{4dmink entropy trunc} how the  SSEE is modified with the application of the various truncations. An area law is recovered with the two number truncations $n_\mx \propto N^{3/4}$, while the linear truncation is more consistent with a volume law. This is similar to what we found in the dS cases we studied above. The area law coefficients in most cases are $\sim 1$ and are therefore close to the expected value $0.78$. What is more challenging and non-trivial here, compared to the dS cases, is achieving complementarity. The geometries of $\diam^4_\ell$ and its complement are very different (see Figure \ref{fig:4ddiamond}) and therefore the truncations ought to take this difference into account.  

As we can see by comparing Figures \ref{4dmink entropy num nofactor} and \ref{4dmink entropy num factor2}, it is correct to truncate the complementary region more than $\diam^4_\ell$. The SSEE can get closer to satisfying both complementarity and the expected area law coefficient if one appropriately tunes the proportionality constant in $n_\mx \propto N^{3/4}$. As discussed in the main text, we do not have a covariant argument by which to uniquely set this proportionality constant. In the absence of such an argument, we do not pursue tuning the constant(s).

\end{appendices}

\bibliography{refs_main}{}
\bibliographystyle{ieeetr} 

\end{document}